\documentclass[notitlepage,a4paper,11pt,amsfonts,amssymb,amsmath]{revtex4-1}
\usepackage{enumerate}
\usepackage{stmaryrd}

\newcommand{\ola}[1]{\overset{{}_{\shortleftarrow}}{#1}}
\newcommand{\ora}[1]{\overset{{}_{\shortrightarrow}}{#1}}

\begin{document}
\title{Parameterized post-Newtonian formalism for multimetric gravity}

\author{Manuel Hohmann}
\email{manuel.hohmann@ut.ee}
\affiliation{Teoreetilise F\"u\"usika Labor, F\"u\"usika Instituut, Tartu \"Ulikool, Riia 142, 51014 Tartu, Estonia}

\begin{abstract}
We discuss the post-Newtonian limit of multimetric gravity theories with \(N \geq 2\) metric tensors  and a corresponding number of standard model copies, and construct an extension of the parameterized post-Newtonian (PPN) formalism. This extended formalism allows a characterization of multimetric gravity theories by a set of constant parameters. The multimetric PPN parameters we derive are a superset of the standard PPN parameters, which have been measured using high-precision experiments in the solar system. We apply our formalism to a class of theories which we previously discussed in the context of cosmology and gravitational waves, and which feature an accelerating expansion of the universe. A comparison between our results and the measured PPN parameters shows that multimetric gravity is fully compatible with solar system observations.
\end{abstract}
\maketitle

\section{Introduction}\label{sec:introduction}
In this article we continue our discussion of multimetric gravity theories with \(N \geq 2\) metric tensors \(g^I_{ab}\) and a corresponding number of standard model copies \(\varphi^I\)~\cite{Hohmann:2009bi,Hohmann:2010ni,Hohmann:2010vt,Hohmann:2011gb}. These theories have been constructed such that each standard model copy couples to only one of the metric tensors, and the interaction between the different standard model copies is mediated only by a coupling of the different metrics. It then follows that each matter type \(\varphi^I\) appears dark to observers constituted by a different matter type \(\varphi^J\), i.e., it cannot be observed by other means than its gravitational interaction. In particular we are interested in theories with a Newtonian limit in which gravity is attractive within each of the matter sectors, but repulsive of equal strength between different standard model copies. While this is not possible in the bimetric case \(N = 2\)~\cite{Hohmann:2009bi}, it leads to an accelerating expansion of the universe that naturally becomes small at late times for \(N \geq 3\) metrics~\cite{Hohmann:2010vt}. Our aim is to further test the predictions of our repulsive gravity models using high-precision experiments in the solar system. For this purpose we here construct an extension of the parameterized post-Newtonian (PPN) formalism~\cite{Nordtvedt:1968qs,Thorne:1970wv,Will:1993ns,Will:2005va,Will:2014kxa} to multimetric gravity theories. In earlier work we have already constructed a simple extension of the PPN formalism to linearized multimetric gravity~\cite{Hohmann:2010ni}. We now extend this formalism to the full, non-linear, post-Newtonian level.

The post-Newtonian limit of several gravity theories with more than one metric tensor has been previously discussed. This includes bimetric theories in which only one metric is dynamical, while the second metric is a fixed background~\cite{Rosen:1973,Rosen:1974ua}, and which possibly include further dynamical gravitational fields, such as vectors~\cite{Rastall:1976uh,Rastall:1977xy} or a rank two tensor which algebraically determines the metric~\cite{Lightman:1973}. Theories of this type have been analyzed using the standard PPN formalism~\cite{Will:1993ns,Lee:1975kc}. A different type of bimetric theories, in which both metrics are dynamical, has become known as bigravity~\cite{Isham:1971gm,Damour:2002ws}. These theories are closely connected to covariant theories of massive gravity, where a second metric is required in order to construct a covariant mass term, and have received recent attention since a class of them has been shown to be ghost free; see~\cite{Boulware:1973my} for the ghost problem of massive gravity and \cite{deRham:2010ik,deRham:2010kj,Hassan:2011vm,Hassan:2011hr,deRham:2011rn,Hassan:2011tf,Hassan:2011zd,deRham:2013tfa} for ghost free theories. The post-Newtonian limit of such theories has been studied in an extended version of the standard PPN formalism~\cite{Clifton:2010hz}. However, these bigravity theories are in general different from the multimetric gravity theories we consider here: they contain only one copy of standard model matter, which couples to one of the metric tensors, or to both metrics as in~\cite{Akrami:2013ffa,Tamanini:2013xia}. In contrast, we here consider theories where each metric governs the dynamics of a different standard model copy. Theories of this type have been studied, e.g., in the bimetric case~\cite{Berezhiani:2007zf,Berezhiani:2008nr}, using a third metric~\cite{Berezhiani:2009kv} or using additional tensor fields that mediate the gravitational interaction between different standard model copies~\cite{Hossenfelder:2008bg}. We do not consider additional fields besides the metrics and matter fields in this article.

A generic property of theories in which the dynamics of different matter types is governed by different metrics is a breaking of the weak equivalence principle, which states that all freely falling test masses follow the same trajectories, independent of their mass and composition. Indeed the universality of free fall is clearly violated if test masses constituted by different matter types follow the geodesics of different metrics. One might therefore argue that theories of these type would be non-viable, since the weak equivalence principle has been shown to hold in high-precision experiments, see e.g.~\cite{Will:2005va,Will:2014kxa,Eotvos:1922pb,Roll:1964rd,Nordtvedt:1968qr,Su:1994gu,Peters:1999,Williams:2005rv} and the recent focus issue Vol. 29, Number 18 of \emph{Classical and Quantum Gravity}~\cite{Speake:2012}. However, these experiments test the weak equivalence principle only for the visible copy of standard model matter, which is observable through its non-gravitational interaction. This opens the possibility that test masses constituted by additional, dark standard model copies follow different trajectories. In this article we therefore consider only theories which obey the experimentally verified universality of free fall for visible matter, and allow different free fall trajectories only for dark test masses, which are not accessible to laboratory experiments.

The PPN formalism, which in its standard form characterizes gravity theories with a single metric by a set of ten parameters, has evolved to an important testbed for gravitational theories. Solar system experiments have placed tight bounds on the PPN parameters, see e.g.~\cite{Will:2005va,Will:2014kxa,Fomalont:2009zg,Bertotti:2003rm,Hofmann:2010,Lambert:2011,Fienga:2011qh,Pitjeva:2013xxa}. Gravity theories whose parameters exceed these bounds are therefore experimentally excluded. This standard version of the PPN formalism can in principle also be applied to multimetric gravity theories of the type we discuss here, in which one metric and one standard model copy govern the dynamics of the solar system. However, it turns out that the metrics in the post-Newtonian limit of multimetric gravity theories may be of a more general form than the standard PPN metric, so that new PPN potentials must be introduced. It is the primary aim of this article to augment the standard PPN formalism with these additional PPN potentials, and correspondingly with additional PPN parameters, whose values are accessible to solar system experiments. Another aim of this article is to further augment the standard PPN formalism to also describe the gravitational interaction between different standard model copies in the post-Newtonian limit. An extension of this type allows, for example, to study the deflection of visible light by dark galaxies. In this article we present an extension of the standard PPN formalism which accomplishes both of these aims. In order to demonstrate the usefulness of this extended formalism we apply it to a generic class of multimetric theories. A comparison of the calculated and measured values of the PPN parameters then enables us to use their bounds as a strong viability test for multimetric gravity.

The outline of this article is as follows. In section~\ref{sec:multimetric} we give a concise definition of the multimetric gravity theories we discuss here. We then derive their post-Newtonian limit in section~\ref{sec:multippn}, and thereby construct an extension of the parameterized post-Newtonian formalism to multimetric gravity. We elaborate on the relation between between our newly developed formalism and the standard PPN formalism in section~\ref{sec:standardppn}. In particular we relate the multimetric PPN parameters to the standard PPN parameters, which are accessible to experiments in the solar system. We explicitly calculate these PPN parameters for a general multimetric gravity theory in section~\ref{sec:calculation}. In section~\ref{sec:applications} we apply this general formalism to two concrete gravity theories, before we end with a conclusion in section~\ref{sec:conclusion}.

\section{Multimetric gravity}\label{sec:multimetric}
The starting point of our construction is a concise definition of the multimetric gravity theories to which our extended version of the PPN formalism will apply. Our primary aim is to apply it to the theories we have discussed in earlier works~\cite{Hohmann:2009bi,Hohmann:2010ni,Hohmann:2010vt,Hohmann:2011gb}. Their properties can be summarized using the following set of assumptions:
\begin{enumerate}[\it (i)]
\item\label{ass:fields}
The field content is given by \(N \geq 2\) copies \(\varphi^1, \ldots, \varphi^N\) of standard model matter and a corresponding number of metric tensors \(g^1_{ab}, \ldots, g^N_{ab}\).

\item\label{ass:action}
The dynamics are governed by a diffeomorphism invariant action of the type
\begin{equation}\label{eqn:actionsplit}
S = S_G[g^1, \ldots, g^N] + \sum_{I = 1}^{N}S_M[g^I, \varphi^I]\,,
\end{equation}
where \(S_M\) denotes the standard model action.

\item\label{ass:tensor}
The field equations are obtained by variation with respect to the metrics \(g^1_{ab}, \ldots, g^N_{ab}\), and so are a set of symmetric two-tensor equations of the form \(K^I_{ab} = 8\pi G_NT^I_{ab}\).

\item\label{ass:derivatives}
The geometry tensor \(K^I_{ab}\) contains at most second derivatives of the metric, which can be achieved by a suitable choice of the gravitational action~\eqref{eqn:actionsplit}.

\item\label{ass:flat}
The vacuum solution is given by a set of flat metrics \(g^I_{ab} = \eta_{ab}\).
\end{enumerate}
Assumption~\textit{(\ref{ass:action})} implies that each standard model copy \(\varphi^I\) couples only to its corresponding metric tensor \(g^I_{ab}\). This ensures that the dynamics and causality of each standard model copy are governed by a single metric. It further ensures that the interaction between the different standard model copies is mediated only through gravity, so that they appear mutually dark. Variation of the gravitational action \(S_G\) with respect to the metrics \(g^I_{ab}\) then yields the geometry tensors \(K^I_{ab}\), while variation of the matter action \(S_M\) yields the usual energy-momentum tensors \(T^I_{ab}\), as stated in assumption~\textit{(\ref{ass:tensor})}. Assumption~\textit{(\ref{ass:derivatives})} is a technical requirement, which we use here in order to restrict the possible terms that may appear in the post-Newtonian limit. We further exclude cosmological constants by assumption~\textit{(\ref{ass:flat})}.

We will make use of these assumptions throughout the following sections. In particular we will discuss which restrictions we obtain on the post-Newtonian limit of multimetric gravity theories. We will derive this limit in the following section.

\section{Multimetric PPN formalism}\label{sec:multippn}
We are now able to derive the post-Newtonian limit of a multimetric gravity theory satisfying assumptions~\textit{(\ref{ass:fields})}--\textit{(\ref{ass:flat})} displayed in the preceding section, and construct an extension to the parameterized post-Newtonian (PPN) formalism detailed in~\cite{Will:1993ns}. The starting point of our construction will be a post-Newtonian expansion of the metrics \(g^I_{ab}\), and thus the geometry side of the gravitational field equations, as shown in section~\ref{subsec:ppnmetric}. A corresponding expansion of the matter side of the field equations will be performed in section~\ref{subsec:matter}. We then elaborate on various properties of the PPN metrics. In particular, we will discuss issues related to gauge invariance in section~\ref{subsec:ppngauge} and calculate its behavior under Lorentz transformations in section~\ref{subsec:lorentz}.

\subsection{Post-Newtonian metric}\label{subsec:ppnmetric}
A basic ingredient of the PPN formalism is a perturbative expansion of the metrics \(g^I_{ab}\) in orders of the velocity \(\vec{v}\) of the source matter in a given frame of reference. Using assumption~\textit{(\ref{ass:flat})} this is a weak field approximation around the flat vacuum metric \(\eta_{ab}\) in Cartesian coordinates \((x^a) = (t, \vec{x})\),
\begin{equation}\label{eqn:perturbation}
g^I_{ab} = \eta_{ab} + h^I_{ab} = \eta_{ab} + h^{I(1)}_{ab} + h^{I(2)}_{ab} + h^{I(3)}_{ab} + h^{I(4)}_{ab}\,,
\end{equation}
where each term \(h^{I(n)}_{ab}\) is of order \(|\vec{v}|^n \equiv \mathcal{O}(n)\). In order to describe the motion of test bodies in the lowest post-Newtonian approximation an expansion up to the fourth velocity order \(\mathcal{O}(4)\) is sufficient. A detailed analysis shows that not all components of the metric perturbations need to be expanded to the fourth velocity order, while others vanish due to Newtonian energy conservation or time reversal symmetry. In the following we list only the relevant, non-vanishing components of the metric perturbations. These are written in terms of the so-called PPN potentials \(\chi^I, W^{I\pm}, \Phi_p^I, \Phi_{\Pi}^I, \Omega_1^I, \Omega_2^I, \Psi_1^{IJ}, \ldots, \Psi_7^{IJ}\) and constant PPN parameters \(\alpha^{IJ}, \gamma^{IJ}, \theta^{IJ}, \sigma_{\pm}^{IJ}, \phi_p^{IJ}, \phi_{\Pi}^{IJ}, \omega_1^{IJ}, \omega_2^{IJ}, \psi_1^{IJK}, \ldots, \psi_7^{IJK}\) as
\begin{subequations}\label{eqn:ppnmetric}
\begin{align}
h^{I(2)}_{00} &= -\sum_{J = 1}^{N}\alpha^{IJ}\triangle\chi^J\,,\label{eqn:h200}\\
h^{I(2)}_{\alpha\beta} &= \sum_{J = 1}^{N}\left(2\theta^{IJ}\chi^J_{,\alpha\beta} - (\gamma^{IJ} + \theta^{IJ})\triangle\chi^J\delta_{\alpha\beta}\right)\,,\label{eqn:h2ab}\\
h^{I(3)}_{0\alpha} &= \sum_{J = 1}^{N}\left(\sigma_+^{IJ}W^{J+}_{\alpha} + \sigma_-^{IJ}W^{J-}_{\alpha}\right)\,,\label{eqn:h30a}\\
h^{I(4)}_{00} &= \sum_{J = 1}^{N}\left(\phi_p^{IJ}\Phi_p^J + \phi_{\Pi}^{IJ}\Phi_{\Pi}^J + \sum_{A = 1}^{2}\omega_A^{IJ}\Omega_A^J\right) + \sum_{J,K = 1}^{N}\sum_{A = 1}^{7}\psi_A^{IJK}\Psi_A^{JK}\,.\label{eqn:h400}
\end{align}
\end{subequations}
The spacetime dependent post-Newtonian potentials appearing in the metric above are Poisson-like integrals over the source matter distribution, which is assumed to be a perfect fluid with rest energy density \(\rho^I\), velocity \(v^I_{\alpha}\), internal energy density \(\rho^I\Pi^I\) and pressure \(p^I\) for each matter type \(I = 1, \ldots, N\). The velocity orders assigned to these quantities are \(\rho^I \sim \mathcal{O}(2)\) and \(\rho^I\Pi^I, p^I \sim \mathcal{O}(4)\), based on their values in the solar system. The potentials are then given by the ``superpotential''
\begin{equation}\label{eqn:chipot}
\chi^I(t,\vec{x}) = -\int\rho^I(t,\vec{x}')|\vec{x} - \vec{x}'|d^3x'\,,
\end{equation}
the vector potentials
\begin{equation}\label{eqn:wpot}
W^{\pm I}_{\alpha}(t,\vec{x}) = \int\rho^I(t,\vec{x}')\left(\frac{v^I_{\alpha}(t,\vec{x}')}{|\vec{x} - \vec{x}'|} \pm \frac{(x_{\alpha} - x_{\alpha}')\left[\vec{v}^I(t,\vec{x}') \cdot (\vec{x} - \vec{x}')\right]}{|\vec{x} - \vec{x}'|^3}\right)d^3x'\,,
\end{equation}
the pressure and internal energy
\begin{equation}\label{eqn:phipot}
\Phi_p^I(t,\vec{x}) = \int\frac{p^I(t,\vec{x}')}{|\vec{x} - \vec{x}'|}d^3x'\,, \qquad
\Phi_{\Pi}^I(t,\vec{x}) = \int\frac{\rho^I(t,\vec{x}')\Pi^I(t,\vec{x}')}{|\vec{x} - \vec{x}'|}d^3x'\,,
\end{equation}
the kinetic energy
\begin{equation}\label{eqn:omegapot}
\Omega_1^{I}(t,\vec{x}) = \int\frac{\rho^I(t,\vec{x}'){v^I}^2(t,\vec{x}')}{|\vec{x} - \vec{x}'|}d^3x'\,, \qquad
\Omega_2^{I}(t,\vec{x}) = \int\frac{\rho^I(t,\vec{x}')\left[\vec{v}^I(t,\vec{x}') \cdot (\vec{x} - \vec{x}')\right]^2}{|\vec{x} - \vec{x}'|^3}d^3x'\,,
\end{equation}
and the non-linear potentials, which can most conveniently be defined by their double Laplacians
\begin{gather}
\triangle\triangle\Psi_1^{IJ} = \triangle\chi^I\triangle\triangle\triangle\chi^J\,, \qquad
\triangle\triangle\Psi_2^{IJ} = \chi^I_{,\alpha\beta}\triangle\triangle\chi^J_{,\alpha\beta}\,, \qquad
\triangle\triangle\Psi_3^{IJ} = \triangle\chi^I_{,\alpha}\triangle\triangle\chi^J_{,\alpha}\,,\nonumber\\
\triangle\triangle\Psi_4^{IJ} = \chi^I_{,\alpha\beta\gamma}\triangle\chi^J_{,\alpha\beta\gamma}\,, \qquad
\triangle\triangle\Psi_5^{IJ} = \triangle\triangle\chi^I\triangle\triangle\chi^J\,,\label{eqn:psipot}\\
\triangle\triangle\Psi_6^{IJ} = \triangle\chi^I_{,\alpha\beta}\triangle\chi^J_{,\alpha\beta}\,, \qquad
\triangle\triangle\Psi_7^{IJ} = \chi^I_{,\alpha\beta\gamma\delta}\chi^J_{,\alpha\beta\gamma\delta}\,.\nonumber
\end{gather}
with \(\triangle = \partial^{\alpha}\partial_{\alpha}\). Here we have chosen units in which the Newtonian gravitational constant takes the value \(G_N = 1\). We further assume that the gravitational field is quasi-static, so that changes are only induced by the motion of the source matter. Time derivatives \(\partial_0\) of all quantities are therefore weighted with an additional velocity order \(\mathcal{O}(1)\).

The PPN parameters \(\alpha^{IJ}, \gamma^{IJ}, \theta^{IJ}, \sigma_{\pm}^{IJ}, \phi_p^{IJ}, \phi_{\Pi}^{IJ}, \omega_1^{IJ}, \omega_2^{IJ}, \psi_1^{IJK}, \ldots, \psi_7^{IJK}\) are characteristic for the concrete multimetric gravity theory under consideration. They can be determined from a perturbative solution of the gravitational field equations. We will provide this solution in section~\ref{sec:calculation}.

\subsection{Matter content}\label{subsec:matter}
As already mentioned in the preceding section we assume the source matter to be a perfect fluid with rest energy density \(\rho^I\), velocity \(v^I_{\alpha}\), internal energy density \(\rho^I\Pi^I\) and pressure \(p^I\) for each matter type \(I = 1, \ldots, N\). Under this assumption the components of the energy-momentum tensors \(T^I_{ab}\), which enter the field equations as stated in assumption~\textit{(\ref{ass:tensor})}, take the form
\begin{subequations}\label{eqn:energymomentum}
\begin{align}
T^I_{00} &= \rho^I\left(1 + \Pi^I + {v^I}^2 + \sum_{J = 1}^{N}\alpha^{IJ}\triangle\chi^J\right) + \mathcal{O}(6)\,,\\
T^I_{0\alpha} &= -\rho^Iv^I_{\alpha} + \mathcal{O}(5)\,,\\
T^I_{\alpha\beta} &= \rho^Iv^I_{\alpha}v^I_{\beta} + p^I\delta_{\alpha\beta} + \mathcal{O}(6)\,.
\end{align}
\end{subequations}
We further note that energy and momentum are covariantly conserved, \(\nabla^I_aT^{I\,ab} = 0\), where \(\nabla^I\) denotes the Levi-Civita connection of \(g^I_{ab}\). This is the case for any type of matter \(\varphi^I\) whose dynamics is governed by a diffeomorphism invariant action \(S_M[g^I, \varphi^I]\), as required by our assumption~\textit{(\ref{ass:action})}. We can decompose the conservation equation into time and space components and insert the perfect fluid energy momentum tensor~\eqref{eqn:energymomentum}. From this we obtain
\begin{subequations}
\begin{align}
0 &= \nabla^I_aT^{I\,a0} = \rho^I_{,0} + (\rho^Iv^I_{\alpha})_{,\alpha} + \mathcal{O}(5)\,,\label{eqn:conservation1}\\
0 &= \nabla^I_aT^{I\,a\alpha} = \rho^I\frac{dv^I_{\alpha}}{dt} + p^I_{,\alpha} + \frac{1}{2}\rho^I\sum_{J = 1}^{N}\alpha^{IJ}\triangle\chi^J_{,\alpha} + \mathcal{O}(6)\,.\label{eqn:conservation2}
\end{align}
\end{subequations}
The first equation is simply the continuity equation, while the second equation corresponds to the Eulerian equation of motion for a perfect fluid, adapted to multimetric gravity. They are helpful for deriving relations between the different PPN potentials. From the continuity equation we obtain
\begin{equation}
\chi^I_{,0\alpha} = W^{-I}_{\alpha}\,, \quad \chi^I_{,00} = \Omega_2^I + \Omega_3^I - \Omega_1^I\,,
\end{equation}
where we have introduced
\begin{equation}\label{eqn:omega3}
\Omega_3^{I}(t,\vec{x}) = \int\frac{\rho^I(t,\vec{x}')(\vec{x} - \vec{x}')}{|\vec{x} - \vec{x}'|} \cdot \frac{d\vec{v}^I(t,\vec{x}')}{dt}d^3x'\,.
\end{equation}
Note that \(\Omega_3^I\) is not a separate PPN potential. Using the Eulerian equation of motion we find
\begin{equation}\label{eqn:omegacont}
\Omega_3^I = \frac{1}{2}\sum_{J = 1}^{N}\alpha^{IJ}\left(\Psi_3^{JI} + \Psi_5^{IJ}\right) - 2\Phi_p^I\,,
\end{equation}
which shows that \(\Omega_3^I\) can be expressed in terms of other PPN potentials. We will make use of these relations in the following sections.

\subsection{Gauge transformations}\label{subsec:ppngauge}
Assumption~\textit{(\ref{ass:tensor})} on the class of multimetric gravity theories we consider in this article states that the field equations are obtained by variation from an action, which is invariant under diffeomorphisms according to assumption~\textit{(\ref{ass:action})}. It thus follows that also the field equations and its solutions are diffeomorphism invariant. Every diffeomorphism is generated by a vector field \(\xi\) and changes tensor fields by their Lie derivatives. For the metric tensors we thus find
\begin{equation}\label{eqn:metricchange}
\delta_{\xi}g^I_{ab} = (\mathcal{L}_{\xi}g^I)_{ab} = 2\nabla^I_{(a}\xi_{b)}\,.
\end{equation}
Since we consider only a particular class of post-Newtonian solutions in this article, which are given by the perturbation ansatz~\eqref{eqn:perturbation} and the metric perturbations~\eqref{eqn:ppnmetric}, we will consider only diffeomorphisms which leave the structure of the perturbation ansatz and the expansion in terms of PPN potentials invariant. This is the case only if the vector field \(\xi\) itself can be written in terms of the PPN potentials. It turns out that the only possible and relevant choice for \(\xi\) is given by
\begin{equation}\label{eqn:gaugevec}
\xi_0 = \sum_{I = 1}^{N}\lambda_1^I\chi^I_{,0}\,, \qquad \xi_{\alpha} = \sum_{I = 1}^{N}\lambda_2^I\chi^I_{,\alpha}
\end{equation}
with \(2N\) constants \(\lambda_1^I, \lambda_2^I\). Inserting this into equation~\eqref{eqn:metricchange} we find the change of the metric components
\begin{subequations}\label{eqn:gaugetrans}
\begin{align}
\delta_{\xi}g^I_{00} &= 2\sum_{J = 1}^{N}\lambda_1^J\left(\Omega_2^J + \Omega_3^J - \Omega_1^J\right) - \sum_{J,K = 1}^{N}\lambda_2^J\alpha^{IK}\left(2\Psi_2^{JK} + \Psi_3^{JK} + \Psi_3^{KJ} + 4\Psi_4^{JK} + 4\Psi_6^{JK}\right)\,,\label{eqn:gaugetrans1}\\
\delta_{\xi}g^I_{0\alpha} &= \sum_{J = 1}^{N}\left(\lambda_1^J + \lambda_2^J\right)W^{-J}_{\alpha}\,,\label{eqn:gaugetrans2}\\
\delta_{\xi}g^I_{\alpha\beta} &= 2\sum_{J = 1}^{N}\lambda_2^J\chi^J_{,\alpha\beta}\,.\label{eqn:gaugetrans3}
\end{align}
\end{subequations}
We can choose a gauge so that certain PPN potentials are eliminated from the metric perturbations~\eqref{eqn:ppnmetric}. Using equation~\eqref{eqn:gaugetrans3} we eliminate the anisotropic term \(\chi^I_{,\alpha\beta}\) from the metric perturbation \(h^{I(2)}_{\alpha\beta}\), thus effectively setting the diagonal elements \(\theta^{II}\) to \(0\). This fixes the constants \(\lambda_2^I\). We further use equation~\eqref{eqn:gaugetrans1} together with the Eulerian equation of motion in the form~\eqref{eqn:omegacont} to eliminate the difference of the potentials \(\Psi_1^{II}\) and \(\Psi_5^{II}\) from \(h^{I(4)}_{00}\), which corresponds to setting the diagonal elements \(\psi_5^{III}\) equal to \(\psi_1^{III}\). This finally fixes the constants \(\lambda_1^I\), so that the gauge is completely determined. The choice of this particular gauge fixing will become apparent when we discuss the standard PPN formalism in section~\ref{sec:standardppn}.

\subsection{Lorentz transformations}\label{subsec:lorentz}
In the previous sections we have used a fixed Cartesian coordinate system \((x^a) = (t, \vec{x})\) in which we expressed the metrics \(g^I_{ab}\) and the post-Newtonian potentials. We will now transform the PPN metric to a coordinate system \((\tilde{x}^a) = (\tilde{t}, \vec{\tilde{x}})\) which is moving with a velocity \(\vec{w}\) relative to the previously used coordinate system. Since we wish to retain the order \(|\vec{v}| \sim \mathcal{O}(1)\) of the velocity of the source matter in the new coordinate system, we assume that the relative velocity is of the same order \(|\vec{w}| \sim \mathcal{O}(1)\). We can then expand the coordinate transform in powers of \(\vec{w}\). Up to the required velocity order it takes the post-Galilean form~\cite{Chandrasekhar:1967}
\begin{subequations}\label{eqn:pgtrans}
\begin{align}
x_{\alpha} &= \tilde{x}_{\alpha} + \left(1 + \frac{1}{2}w^2\right)w_{\alpha}\tilde{t} + \frac{1}{2}w_{\alpha}w_{\beta}\tilde{x}_{\beta} + \mathcal{O}(4) \cdot \tilde{x}_{\alpha}\,,\\
t &= \left(1 + \frac{1}{2}w^2 + \frac{3}{8}w^4\right)\tilde{t} + \left(1 + \frac{1}{2}w^2\right)w_{\alpha}\tilde{x}_{\alpha} + \mathcal{O}(5) \cdot \tilde{t}\,.
\end{align}
\end{subequations}
We can then calculate the metric tensors \(\tilde{g}^I_{ab}\) in the moving coordinate system using the standard formula
\begin{equation}\label{eqn:transform}
\tilde{g}^I_{ab} = \frac{dx^i}{d\tilde{x}^a}\frac{dx^j}{d\tilde{x}^b}g^I_{ij}\,.
\end{equation}
In the metric \(\tilde{g}^I_{ab}\) obtained from this calculation we need to express the post-Newtonian potentials \(\chi^I, W^{\pm I}_{\alpha}, \ldots\) in terms of the equivalent expressions \(\tilde{\chi}^I, \tilde{W}^{\pm I}_{\alpha}, \ldots\) in moving coordinates, which are obtained from the definitions in section~\ref{subsec:ppnmetric} by replacing coordinates \(x^a\) with coordinates \(\tilde{x}^a\) and the source matter velocity \(\vec{v}^I\) with the velocity
\begin{equation}
\tilde{v}^I_{\alpha} = v^I_{\alpha} - w_{\alpha} + \mathcal{O}(3)
\end{equation}
in the moving coordinate system. By explicit calculation we obtain the PPN potentials up to the required velocity orders
\begin{subequations}
\begin{align}
\triangle\chi^I &= \triangle\tilde{\chi}^I + 2w_{\alpha}\tilde{W}^{-I}_{\alpha} - w_{\alpha}w_{\beta}\tilde{\chi}^I_{,\alpha\beta} + \mathcal{O}(6)\,,\\
\chi^I_{,\alpha\beta} &= \tilde{\chi}^I_{,\alpha\beta} + \mathcal{O}(4)\,,\\
W^{+I}_{\alpha} &= \tilde{W}^{+I}_{\alpha} + w_{\beta}\tilde{\chi}^I_{,\alpha\beta} - w_{\alpha}\triangle\tilde{\chi}^I + \mathcal{O}(5)\,,\\
W^{-I}_{\alpha} &= \tilde{W}^{-I}_{\alpha} - w_{\beta}\tilde{\chi}^I_{,\alpha\beta} + \mathcal{O}(5)\,,\\
\Omega_1^I &= \tilde{\Omega}_1^I + w_{\alpha}\left(\tilde{W}^{+I}_{\alpha} + \tilde{W}^{-I}_{\alpha}\right) - \frac{1}{2}w^2\triangle\tilde{\chi}^I + \mathcal{O}(6)\,,\\
\Omega_2^I &= \tilde{\Omega}_2^I + w_{\alpha}\left(\tilde{W}^{+I}_{\alpha} - \tilde{W}^{-I}_{\alpha}\right) + w_{\alpha}w_{\beta}\tilde{\chi}^I_{,\alpha\beta} - \frac{1}{2}w^2\triangle\tilde{\chi}^I + \mathcal{O}(6)\,,\\
\Phi_p^I &= \tilde{\Phi}_p^I + \mathcal{O}(6)\,, \quad \Phi_{\Pi}^I = \tilde{\Phi}_{\Pi}^I + \mathcal{O}(6)\,, \quad \Psi_A^{IJ} = \tilde{\Psi}_A^{IJ} + \mathcal{O}(6)\,.
\end{align}
\end{subequations}
Inserting these into the transformation formula~\eqref{eqn:transform} for the metric tensors and decomposing the components of \(\tilde{g}^I_{ab}\) into velocity orders we finally find
\begin{subequations}\label{eqn:ltmetric}
\begin{align}
\tilde{h}^{I(2)}_{00} &= -\sum_{J = 1}^{N}\alpha^{IJ}\triangle\tilde{\chi}^J\,,\\
\tilde{h}^{I(2)}_{\alpha\beta} &= \sum_{J = 1}^{N}\left(2\theta^{IJ}\tilde{\chi}^J_{,\alpha\beta} - (\gamma^{IJ} + \theta^{IJ})\triangle\tilde{\chi}^J\delta_{\alpha\beta}\right)\,,\\
\tilde{h}^{I(3)}_{0\alpha} &= \sum_{J = 1}^{N}\left(\sigma_+^{IJ}\tilde{W}^{J+}_{\alpha} + \sigma_-^{IJ}\tilde{W}^{J-}_{\alpha} - (\alpha^{IJ} + \gamma^{IJ} + \theta^{IJ} + \sigma_+^{IJ})w_{\alpha}\triangle\tilde{\chi}^J + (2\theta^{IJ} + \sigma_+^{IJ} - \sigma_-^{IJ})w_{\beta}\tilde{\chi}^J_{,\alpha\beta}\right)\,,\\
\tilde{h}^{I(4)}_{00} &= \sum_{J = 1}^{N}\Bigg(\phi_p^{IJ}\tilde{\Phi}_p^J + \phi_{\Pi}^{IJ}\tilde{\Phi}_{\Pi}^J + \sum_{A = 1}^{2}\omega_A^{IJ}\tilde{\Omega}_A^J + (2\sigma_-^{IJ} + \omega_1^{IJ} - \omega_2^{IJ} - 2\alpha^{IJ})w_{\alpha}\tilde{W}^{-J}_{\alpha}\nonumber\\
&\phantom{=}+ (2\sigma_+^{IJ} + \omega_1^{IJ} + \omega_2^{IJ})w_{\alpha}\tilde{W}^{+J}_{\alpha} - \left(\alpha^{IJ} + \gamma^{IJ} + \theta^{IJ} + 2\sigma_+^{IJ} + \frac{1}{2}\omega_1^{IJ} + \frac{1}{2}\omega_2^{IJ}\right)w^2\triangle\tilde{\chi}^J\\
&\phantom{=}+ (\alpha^{IJ} + 2\theta^{IJ} + 2\sigma_+^{IJ} - 2\sigma_-^{IJ} + \omega_2^{IJ})w_{\alpha}w_{\beta}\tilde{\chi}^J_{,\alpha\beta}\Bigg) + \sum_{J,K = 1}^{N}\sum_{A = 1}^{7}\psi_A^{IJK}\tilde{\Psi}_A^{JK}\,.\nonumber
\end{align}
\end{subequations}
In addition to the PPN potentials defined in section~\ref{subsec:ppnmetric} the metric now contains terms that explicitly depend on the velocity \(\vec{w}\). This more general metric is form invariant under a larger class of diffeomorphisms than we discussed already in section~\ref{subsec:ppngauge}. In addition to the the gauge transform generated by the vector field~\eqref{eqn:gaugevec} we may now consider the vector field \(\xi'\) defined by
\begin{equation}\label{eqn:gaugevec2}
\xi'_0 = \sum_{I = 1}^{N}\lambda_3^Iw_{\alpha}\chi^I_{,\alpha}\,, \qquad \xi'_{\alpha} = 0\,.
\end{equation}
Under this gauge transform the metrics change by
\begin{equation}
\delta_{\xi'}\tilde{g}^I_{00} = 2\sum_{J = 1}^{N}\lambda_3^Jw_{\alpha}W^{-J}_{\alpha}\,, \quad
\delta_{\xi'}\tilde{g}^I_{0\alpha} = \sum_{J = 1}^{N}\lambda_3^Jw_{\beta}\chi^J_{,\alpha\beta}\,, \quad
\delta_{\xi'}\tilde{g}^I_{\alpha\beta} = 0\,.
\end{equation}
We use this additional gauge freedom to eliminate the term \(w_{\beta}\tilde{\chi}^I_{,\alpha\beta}\) from the metric component \(\tilde{h}^{I(3)}_{0\alpha}\), i.e., we apply the aforementioned gauge transform with
\begin{equation}
\lambda_3^I = -(2\theta^{II} + \sigma_+^{II} - \sigma_-^{II})\,.
\end{equation}
The remaining terms involving the velocity \(\vec{w}\) cannot be eliminated by post-Newtonian gauge transformations. Their presence in the PPN metric~\eqref{eqn:ltmetric} indicates the presence of preferred-frame effects in a gravity theory, i.e., effects which depend on the velocity of the complete experimental setup relative to a preferred frame fixed by, for example, symmetry breaking gravitational background fields in the theory's vacuum solution. Conversely, a theory is free of preferred-frame effects if there are no terms in the gauge fixed PPN metric that depend on \(\vec{w}\). This is the case if and only if the PPN parameters satisfy
\begin{subequations}\label{eqn:lorentzinv}
\begin{align}
\alpha^{IJ} + \gamma^{IJ} + \theta^{IJ} + \sigma_+^{IJ} &= 0\,,\\
2\sigma_+^{IJ} + \omega_1^{IJ} + \omega_2^{IJ} &= 0\,,\\
\alpha^{IJ} + 2\theta^{IJ} - 2\sigma_-^{IJ} - \omega_1^{IJ} &= 0\,,\\
2\theta^{IJ} + \sigma_+^{IJ} - \sigma_-^{IJ} - 2\theta^{II} - \sigma_+^{II} + \sigma_-^{II} &= 0
\end{align}
\end{subequations}
for all \(I,J = 1, \ldots, N\). Since there is no experimental evidence for preferred-frame effects, we are particularly interested in gravity theories which satisfy these constraints. Note that this absence of preferred-frame effects is similarly represented in the standard PPN formalism as the condition \(\alpha_1 = \alpha_2 = \alpha_3 = 0\) on three PPN parameters, which is derived in full analogy to the calculation shown in this section~\cite{Thorne:1970wv,Will:1993ns}.

This concludes our construction of a parameterized post-Newtonian formalism for multimetric gravity theories. In the following section we will discuss its relation to the standard PPN formalism for gravity theories with a single metric tensor, before we present a general method for calculating the multimetric PPN parameters in section~\ref{sec:calculation}.

\section{Relation to the standard PPN formalism}\label{sec:standardppn}
The construction presented in the previous section provides an extension to the well-known parameterized post-Newtonian formalism, which has become a useful tool for testing the viability of alternative gravity theories, see~\cite{Will:1993ns} for a review. We will now sketch the basic ingredients of this standard PPN formalism. Since we are mainly interested in experimental tests of gravity theories, we focus on the PPN parameters, which have been measured with high precision in the solar system. In order to make use of these experiments for testing multimetric gravity theories, we derive the relation between these standard PPN parameters and their multimetric counterparts discussed in the preceding section. This connection provides us with a simple viability test for multimetric gravity.

In its most widely used form the PPN formalism is applicable to gravity theories in which a single metric \(g_{ab}\) governs the geodesic motion of test particles. This metric is expanded in analogy to the multimetric perturbation ansatz~\eqref{eqn:perturbation}, where the non-vanishing terms in the perturbative expansion are given by
\begin{subequations}\label{eqn:standardppn}
\begin{align}
h^{(2)}_{00} &= 2\alpha U\,,\\
h^{(2)}_{\alpha\beta} &= 2\gamma U\delta_{\alpha\beta}\,,\\
h^{(3)}_{0\alpha} &= -\frac{1}{2}(3 + 4\gamma + \alpha_1 - \alpha_2 + \zeta_1 - 2\xi)V_{\alpha} - \frac{1}{2}(1 + \alpha_2 - \zeta_1 + 2\xi)W_{\alpha}\,,\\
h^{(4)}_{00} &= -2\beta U^2 - 2\xi\Phi_W + (2 + 2\gamma + \alpha_3 + \zeta_1 - 2\xi)\Phi_1 + 2(1 + 3\gamma - 2\beta + \zeta_2 + \xi)\Phi_2\\
&\phantom{=}+ 2(1 + \zeta_3)\Phi_3 + 2(3\gamma + 3\zeta_4 - 2\xi)\Phi_4 - (\zeta_1 - 2\xi)\mathcal{A}\nonumber
\end{align}
\end{subequations}
in the standard PPN gauge. The PPN potentials \(U, V_{\alpha}, W_{\alpha}, \Phi_W, \Phi_1, \ldots, \Phi_4, \mathcal{A}\) are determined by the matter source of gravity, which is assumed to be a single perfect fluid with density \(\rho\), pressure \(p\), specific internal energy \(\Pi\) and velocity \(\vec{v}\). In terms of these quantities the second order potential \(U\) is given by the Newtonian potential
\begin{equation}
U(t,\vec{x}) = \int d^3x'\frac{\rho(t,\vec{x}')}{|\vec{x} - \vec{x}'|}\,,
\end{equation}
while the third order potentials \(V\) and \(W\) are given by
\begin{subequations}
\begin{align}
V_{\alpha}(t,\vec{x}) &= \int d^3x'\frac{\rho(t,\vec{x}')v_{\alpha}(t,\vec{x}')}{|\vec{x} - \vec{x}'|}\,,\\
W_{\alpha}(t,\vec{x}) &= \int d^3x'\frac{\rho(t,\vec{x}')v_{\beta}(t,\vec{x}')(x_{\alpha} - x_{\alpha}')(x_{\beta} - x_{\beta}')}{|\vec{x} - \vec{x}'|^3}\,.
\end{align}
\end{subequations}
Similar expressions define the fourth order potentials \(\Phi_W, \Phi_1, \ldots, \Phi_4, \mathcal{A}\), as displayed in~\cite{Will:1993ns}. The constant \(\alpha\) corresponds to the effective gravitational constant and is conventionally set to \(1\) by an appropriate choice of units. The remaining constants \(\beta, \gamma, \alpha_1, \ldots, \alpha_3, \zeta_1, \ldots, \zeta_4, \xi\) are the PPN parameters, which can be determined both theoretically and experimentally, thus providing a test of the theory under consideration. Measurements in the solar system indicate that they take the values \(\beta = \gamma = 1\), while all other parameters vanish.

In order to use these measurements as a test for multimetric gravity theories we need to compare the multimetric PPN parameters appearing in the metric perturbations~\eqref{eqn:ppnmetric} with the measured PPN parameters listed above. For this purpose we need to derive an effective theory for a single perfect fluid matter source and a single metric governing the geodesic motion of test masses. In the solar system both gravitational matter sources and test masses are constituted by visible matter, which we choose to identify with the first of the standard model copies \(\varphi^1\) introduced in assumption~\textit{(\ref{ass:fields})}. For the gravitational matter source we thus identify the matter variables \({\rho = \rho^1}\), \({p = p^1}\), \({\Pi = \Pi^1}\) and \({\vec{v} = \vec{v}^1}\). We assume that this is the only matter source within the solar system, so that \({T^2_{ab} = \ldots = T^N_{ab} = 0}\). From the structure of the action displayed in assumption~\textit{(\ref{ass:action})} it follows that the motion of test masses constituted by the same standard model copy \(\varphi^1\) is governed exclusively by the metric \(g^1_{ab}\). In our effective theory we therefore identify the single metric \({g_{ab} = g^1_{ab}}\).

Using the identifications of the matter variables listed above we can express the PPN potentials in the metric perturbations~\eqref{eqn:standardppn} in terms of the multimetric PPN potentials displayed in section~\ref{subsec:ppnmetric} in the form
\begin{gather}
U = -\frac{1}{2}\triangle\chi^1\,, \quad V_{\alpha} = \frac{W^{+1}_{\alpha} + W^{-1}_{\alpha}}{2}\,, \quad W_{\alpha} = \frac{W^{+1}_{\alpha} - W^{-1}_{\alpha}}{2}\,, \quad \mathcal{A} = \Omega_2^1\,, \quad \Phi_1 = \Omega_1^1\,,\nonumber\\
\Phi_2 = \frac{1}{4}\Psi_1^{11} + \frac{1}{2}\Psi_3^{11} + \frac{1}{4}\Psi_5^{11}\,, \quad \Phi_3 = \Phi_{\Pi}^1\,, \quad \Phi_4 = \Phi_p^1\,,\label{eqn:standardpots}\\
\Phi_W = -\frac{1}{4}\Psi_1^{11} - \Psi_2^{11} - \frac{5}{2}\Psi_3^{11} - 2\Psi_4^{11} - \frac{1}{4}\Psi_5^{11} - 3\Psi_6^{11}\,, \quad U^2 = \frac{1}{2}\Psi_1^{11} + 2\Psi_3^{11} + \frac{1}{2}\Psi_5^{11} + \Psi_6^{11}\,.\nonumber
\end{gather}
The converse, however, is not possible, since the non-linear potentials \(\Psi_A^{11}\) cannot be expressed in terms on the standard PPN potentials. The metric perturbations \(h^1_{ab}\) in the multimetric PPN formalism are thus more general than the standard PPN metric perturbations \(h_{ab}\) displayed in equation~\eqref{eqn:standardppn}. In order to compare the two formalisms we thus restrict ourselves to the simpler case in which only the potentials~\eqref{eqn:standardpots} appear in the metric. Setting \({h^1_{ab} = h_{ab}}\) we can then read off the multimetric PPN parameters
\begin{gather}
\alpha^{11} = \alpha\,, \quad \gamma^{11} = \gamma\,, \quad \sigma_+^{11} = -1 - \gamma - \frac{1}{4}\alpha_1\,, \quad \sigma_-^{11} = -\frac{1}{2} - \gamma - \frac{1}{4}\alpha_1 + \frac{1}{2}\alpha_2 - \frac{1}{2}\zeta_1 + \xi\,,\nonumber\\
\phi_{\Pi}^{11} = 2 + 2\zeta_3\,, \quad \phi_p^{11} = 6\gamma + 6\zeta_4 + 4\xi\,, \quad \omega_1^{11} = 2 + 2\gamma + \alpha_3 + \zeta_1 - 2\xi\,, \quad \omega_2^{11} = 2\xi - \zeta_1\,,\nonumber\\
\psi_1^{111} = \psi_5^{111} = \frac{1}{2} + \frac{3}{2}\gamma - 2\beta + \frac{1}{2}\zeta_2 + \xi\,, \quad \psi_2^{111} = 2\xi\,, \quad \psi_3^{111} = 1 + 3\gamma - 6\beta + \zeta_2 + 6\xi\,,\label{eqn:standardparams}\\
\psi_4^{111} = 4\xi\,, \quad \psi_6^{111} = 6\xi - 2\beta\,, \quad \theta^{11} = \psi_7^{111} = 0\,.\nonumber
\end{gather}
Note that this result is compatible with the gauge fixing \(\theta^{II} = 0\) and \(\psi_1^{III} = \psi_5^{III}\) we have chosen in section~\ref{subsec:ppngauge}. This compatibility is the reason for our gauge choice. Inserting the measured values of the standard PPN parameters then directly yields us the expected values of the multimetric PPN parameters
\begin{gather}
\alpha^{11} = \gamma^{11} = 1\,, \quad \sigma_+^{11} = \psi_3^{111} = \psi_6^{111} = -2\,, \quad \sigma_-^{11} = -\frac{3}{2}\,, \quad \phi_{\Pi}^{11} = 2\,, \quad \phi_p^{11} = 6\,,\nonumber\\
\omega_1^{11} = 4\,, \quad \theta^{11} = \omega_2^{11} = \psi_1^{111} = \psi_2^{111} = \psi_4^{111} = \psi_5^{111} = \psi_7^{111} = 0\,,\label{eqn:measured}
\end{gather}
where the values of \(\theta^{11}\) and \(\psi_5^{111}\) are fixed through gauge conditions, and \(\alpha^{11}\) can always be rescaled to \(1\) through a suitable choice of units. This leaves us with thirteen physical PPN parameters in the visible sector. Any concrete multimetric gravity theory with these values of the PPN parameters is compatible with observations of post-Newtonian physics in the solar system. However, it is also possible that theories which yield other values are compatible with experiments. This is due to the fact that only the ten standard PPN parameters have been measured. The newly introduced multimetric PPN parameters may therefore correspond to effects which are not visible to current experiments. A detailed analysis of current and possible future experiments is necessary in order to determine the remaining bounds on the multimetric PPN parameters.

We will make use of the experimental consistency conditions~\eqref{eqn:measured} when we discuss the viability of two example theories in section~\ref{sec:applications}.

\section{Calculation of PPN parameters}\label{sec:calculation}
We will now calculate the PPN parameters in the metric perturbations~\eqref{eqn:ppnmetric} for a general multimetric gravity theory. For this purpose we will perturbatively solve the gravitational field equations \({K^I_{ab} = 8\pi T^I_{ab}}\), which govern the dynamics of multimetric gravity according to assumption~\textit{(\ref{ass:tensor})} stated in section~\ref{sec:multimetric}. In addition to the gravitational field equations we will make use of the gauge fixing detailed in section~\ref{subsec:ppngauge}, which is necessary due to the diffeomorphism invariance we stated in assumption~\textit{(\ref{ass:action})}. It is a virtue of the PPN formalism that at each velocity order \(\mathcal{O}(n)\) the field equations are linear in the unknown metric perturbations \(h^{I(n)}_{ab}\), so that they can be solved separately for each of these perturbations. However, each of these solutions will depend on the lower order perturbations of all other metrics due to their mutual interaction. Our calculation proceeds in three steps. In section~\ref{subsec:ppno2} we solve the field equations up to the second velocity order \(\mathcal{O}(2)\) and determine the metric perturbations \(h^{I(2)}_{00}\) and \(h^{I(2)}_{\alpha\beta}\). Using these results we then determine the third order solution \(h^{I(3)}_{0\alpha}\) in section~\ref{subsec:ppno3} and finally the fourth order solution \(h^{I(4)}_{00}\) in section~\ref{subsec:ppno4}. The general result we obtain here will then be applied to concrete gravity theories in the following section~\ref{sec:applications}.

\subsection{Second velocity order $\mathcal{O}(2)$}\label{subsec:ppno2}
In the first step of our calculation we determine the metric perturbation up to the second velocity order \(\mathcal{O}(2)\). The non-vanishing components of the metric perturbation that we need to consider in this step are \(h^{I(2)}_{00}\) and \(h^{I(2)}_{\alpha\beta}\), which involve the PPN parameters \(\alpha^{IJ}, \gamma^{IJ}, \theta^{IJ}\), as can be read off from equation~\eqref{eqn:ppnmetric}. The equations we need to solve are the second order field equations
\begin{subequations}\label{eqn:o2eom}
\begin{align}
K^{I(2)}_{00} &= 8\pi T^{I(2)}_{00}\,,\\
K^{I(2)}_{\alpha\beta} &= 8\pi T^{I(2)}_{\alpha\beta}\,.
\end{align}
\end{subequations}
In order to solve these equations we need to express the components \(K^{I(2)}_{00}\) and \(K^{I(2)}_{\alpha\beta}\) of the curvature tensor in terms of the metric perturbation. It can be shown that their most general form for a multimetric gravity theory compatible with our assumptions~\textit{(\ref{ass:fields})}--\textit{(\ref{ass:flat})} is given by
\begin{subequations}\label{eqn:o2curvature}
\begin{align}
K^{I(2)}_{00} &= \sum_{J = 1}^{N}\left(S^{IJ}_{1}h^{J(2)}_{00,\alpha\alpha} + S^{IJ}_{2}h^{J(2)}_{\alpha\alpha,\beta\beta} + S^{IJ}_{3}h^{J(2)}_{\alpha\beta,\alpha\beta}\right)\,,\\
K^{I(2)}_{\alpha\beta} &= \sum_{J = 1}^{N}\left(C^{IJ}_{1}h^{J(2)}_{00,\gamma\gamma} + C^{IJ}_{2}h^{J(2)}_{\gamma\gamma,\delta\delta} + C^{IJ}_{3}h^{J(2)}_{\gamma\delta,\gamma\delta}\right)\delta_{\alpha\beta}\nonumber\\
&\phantom{=}+ \sum_{J = 1}^{N}\left(T^{IJ}_{1}h^{J(2)}_{00,\alpha\beta} + T^{IJ}_{2}h^{J(2)}_{\gamma\gamma,\alpha\beta} + T^{IJ}_{3}h^{J(2)}_{\alpha\beta,\gamma\gamma} + T^{IJ}_{4}h^{J(2)}_{\gamma(\alpha,\beta)\gamma}\right)\,,
\end{align}
\end{subequations}
where the constants \(S^{IJ}_{1}, \ldots, S^{IJ}_{3}, C^{IJ}_{1}, \ldots, C^{IJ}_{3}, T^{IJ}_{1}, \ldots, T^{IJ}_{4}\) are uniquely determined by the concrete theory under consideration and can be calculated from the linearized field equations. Inserting the PPN metric~\eqref{eqn:ppnmetric} we further derive
\begin{subequations}
\begin{align}
K^{I(2)}_{00} &= \sum_{J = 1}^{N}X^{IJ}_{1}\triangle\triangle\chi^J\,,\\
K^{I(2)}_{\alpha\beta} &= \sum_{J = 1}^{N}\left(X^{IJ}_{2}\triangle\triangle\chi^J\delta_{\alpha\beta} + X^{IJ}_{3}\triangle\chi^J_{,\alpha\beta}\right)\,,
\end{align}
\end{subequations}
where the constants \(X^{IJ}_{1}, \ldots, X^{IJ}_{3}\) are given by
\begin{subequations}\label{eqn:coeffx123}
\begin{align}
X^{IJ}_{1} &= \sum_{K = 1}^{N}\left(-S^{IK}_{1}\alpha^{KJ} - S^{IK}_{2}(3\gamma^{KJ} + \theta^{KJ}) - S^{IK}_{3}(\gamma^{KJ} - \theta^{KJ})\right)\,,\\
X^{IJ}_{2} &= \sum_{K = 1}^{N}\left(-C^{IK}_{1}\alpha^{KJ} - C^{IK}_{2}(3\gamma^{KJ} + \theta^{KJ}) - C^{IK}_{3}(\gamma^{KJ} - \theta^{KJ}) - T^{IK}_{3}(\gamma^{KJ} + \theta^{KJ})\right)\,,\\
X^{IJ}_{3} &= \sum_{K = 1}^{N}\left(-T^{IK}_{1}\alpha^{KJ} - T^{IK}_{2}(3\gamma^{KJ} + \theta^{KJ}) + 2T^{IK}_{3}\theta^{KJ} - T^{IK}_{4}(\gamma^{KJ} - \theta^{KJ})\right)\,.
\end{align}
\end{subequations}
We now have an expression for the geometry side of the field equations~\eqref{eqn:o2eom} in terms of the PPN potentials and PPN parameters displayed in section~\ref{subsec:ppnmetric}. In the next step we consider the matter side of the field equations. Up to the second velocity order the energy-momentum tensor~\eqref{eqn:energymomentum} takes the form
\begin{subequations}
\begin{align}
T^{I(2)}_{00} &= \rho^I\,,\\
T^{I(2)}_{\alpha\beta} &= 0\,.
\end{align}
\end{subequations}
The relation between the matter density \(\rho^I\) and the PPN potential \(\chi^I\) appearing in the field equations at second velocity order is given by the definition~\eqref{eqn:chipot}, which can be written in differential form as
\begin{equation}
\triangle\triangle\chi^I = 8\pi\rho^I\,.
\end{equation}
With this relation the field equations at the second velocity order are now completely expressed in terms of PPN potentials and take the form
\begin{subequations}
\begin{align}
\sum_{J = 1}^{N}X^{IJ}_{1}\triangle\triangle\chi^J &= \triangle\triangle\chi^I\,,\\
\sum_{J = 1}^{N}\left(X^{IJ}_{2}\triangle\triangle\chi^J\delta_{\alpha\beta} + X^{IJ}_{3}\triangle\chi^J_{,\alpha\beta}\right) &= 0\,.
\end{align}
\end{subequations}
From the requirement that they are satisfied for arbitrary matter distributions, and thus arbitrary superpotentials \(\chi^I\), we can immediately read off from the first equation that the coefficient \(X^{IJ}_{1}\) must satisfy
\begin{equation}\label{eqn:eomx1}
X^{IJ}_{1} = \delta^{IJ}\,.
\end{equation}
From the second equation we see that \(\triangle\triangle\chi^J\delta_{\alpha\beta}\) is a pure trace term, while \(\triangle\chi^J_{,\alpha\beta}\) contains both a trace and a trace-free part. In order for this equation to be satisfied thus both coefficients must vanish independently,
\begin{equation}\label{eqn:eomx23}
X^{IJ}_{2} = X^{IJ}_{3} = 0\,.
\end{equation}
Equations~\eqref{eqn:eomx1} and~\eqref{eqn:eomx23} are a set of \(3N^2\) linear equations for the \(3N^2\) PPN parameters \(\alpha^{IJ}, \gamma^{IJ}, \theta^{IJ}\). However, it turns out that these linear equations are not independent. This is a consequence of assumption~\textit{(\ref{ass:action})}, which states that the field equations are determined by a diffeomorphism invariant action, and further restricts the constants appearing in the expansion~\eqref{eqn:o2curvature} of the curvature tensors. These restrictions are the origin of the gauge freedom we discussed in section~\ref{subsec:ppngauge}. Using the gauge condition \(\theta^{II} = 0\) discussed in the same section we finally obtain a full set of linear equations from which we can determine the PPN parameters \(\alpha^{IJ}, \gamma^{IJ}, \theta^{IJ}\).

\subsection{Third velocity order $\mathcal{O}(3)$}\label{subsec:ppno3}
We will now use the result from the preceding section and determine the metric perturbation up to the third velocity order \(\mathcal{O}(3)\). The only unknown, non-vanishing component of the metric perturbation that we need to consider here is \(h^{I(3)}_{0\alpha}\), which involves the PPN parameters \(\sigma_{\pm}^{IJ}\). In order to determine these parameters we need to solve the third order field equations
\begin{equation}\label{eqn:o3eom}
K^{I(3)}_{0\alpha} = 8\pi T^{I(3)}_{0\alpha}\,.
\end{equation}
We can proceed in full analogy to our solution of the second order field equations shown in the preceding section. First we consider the most general form of the third order curvature tensor \(K^{I(3)}_{0\alpha}\), which is given by
\begin{equation}\label{eqn:o3curvature}
K^{I(3)}_{0\alpha} = \sum_{J = 1}^{N}\left(V^{IJ}_{1}h^{J(3)}_{0\alpha,\beta\beta} + V^{IJ}_{2}h^{J(3)}_{0\beta,\alpha\beta} + V^{IJ}_{3}h^{J(2)}_{00,0\alpha} + V^{IJ}_{4}h^{J(2)}_{\beta\beta,0\alpha} + V^{IJ}_{5}h^{J(2)}_{\alpha\beta,0\beta}\right)\,.
\end{equation}
Note that each metric perturbation \(h^{I(3)}_{0\alpha}\) at the third velocity order depends on all metric perturbations \(h^{J(2)}_{00}\) and \(h^{J(2)}_{\alpha\beta}\) at the second velocity order. The constants \(V^{IJ}_{1}, \ldots, V^{IJ}_{5}\) are determined by the linearized field equations of the concrete multimetric gravity theory under consideration. Inserting the PPN metric~\eqref{eqn:ppnmetric} into this equation we can write the curvature tensor in terms of the PPN potentials \(W^{\pm I}_{\alpha}\) in the form
\begin{equation}
K^{I(3)}_{0\alpha} = \sum_{J = 1}^{N}\left(X^{IJ}_{4}\triangle W^{+J}_{\alpha} + X^{IJ}_{5}\triangle W^{-J}_{\alpha}\right)\,,
\end{equation}
where the constants \(X^{IJ}_{4}, X^{IJ}_{5}\) are given by
\begin{subequations}\label{eqn:coeffx45}
\begin{align}
X^{IJ}_{4} &= \sum_{K = 1}^{N}V^{IK}_{1}\sigma_+^{KJ}\,,\\
X^{IJ}_{5} &= \sum_{K = 1}^{N}\left(V^{IK}_{1}\sigma_-^{KJ} + V^{IK}_{2}\sigma_-^{KJ} - V^{IK}_{3}\alpha^{KJ} - V^{IK}_{4}(3\gamma^{KJ} + \theta^{KJ}) - V^{IK}_{5}(\gamma^{KJ} - \theta^{KJ})\right)\,.
\end{align}
\end{subequations}
We also expand the matter side of the field equations~\eqref{eqn:o3eom}. From the energy-momentum tensor~\eqref{eqn:energymomentum} we can read off the third velocity order component
\begin{equation}
T^{I(3)}_{0\alpha} = -\rho^Iv^I_{\alpha}\,.
\end{equation}
This expression is related to the PPN potentials \(W^{\pm I}_{\alpha}\) via their definition~\eqref{eqn:wpot}, from which one finds
\begin{equation}
\triangle W^{+I}_{\alpha} + \triangle W^{-I}_{\alpha} = -8\pi\rho^Iv^I_{\alpha}\,.
\end{equation}
In terms of the PPN potentials the field equations at the third velocity order finally take the form
\begin{equation}
\sum_{J = 1}^{N}\left(X^{IJ}_{4}\triangle W^{+J}_{\alpha} + X^{IJ}_{5}\triangle W^{-J}_{\alpha}\right) = \triangle W^{+I}_{\alpha} + \triangle W^{-I}_{\alpha}\,.
\end{equation}
As already for the second order field equations we require that these are satisfied for arbitrary matter distributions, which determine the PPN potentials \(W^{\pm I}_{\alpha}\). Note that \(W^{+I}_{\alpha}\) is a divergence-free vector, \(W^{+I}_{\alpha,\alpha} = 0\), while \(W^{-I}_{\alpha}\) is a pure divergence, \(W^{-I}_{\alpha} = \chi^I_{,0\alpha}\). The equations of motion thus split into a pure divergence and a divergence-free part. It then follows that the coefficients in these equations must independently satisfy
\begin{equation}
X^{IJ}_{4} = X^{IJ}_{5} = \delta^{IJ}\,.
\end{equation}
The first part yields \(N^2\) linear equations which we can solve for the \(N^2\) parameters \(\sigma_+^{IJ}\). Similarly, the second part yields \(N^2\) linear equations for the \(N^2\) parameters \(\sigma_-^{IJ}\), but it turns out that these are not independent. This is again a consequence of the diffeomorphism invariance that we demand in assumption~\textit{(\ref{ass:action})}, and which requires us to choose a gauge fixing as derived in section~\ref{subsec:ppngauge}. Since our gauge choice involves fourth order potentials, we will defer this issue to the following section, in which we solve the field equations at the fourth velocity order.

\subsection{Fourth velocity order $\mathcal{O}(4)$}\label{subsec:ppno4}
We now come to the final and most involved part of our calculation, which is solving the field equations at the fourth velocity order. The component of the metric perturbation we will determine here is \(h^{I(4)}_{00}\), which will yield us the remaining PPN parameters \(\phi_p^{IJ}, \phi_{\Pi}^{IJ}, \omega_1^{IJ}, \omega_2^{IJ}, \psi_1^{IJK}, \ldots, \psi_7^{IJK}\). It will further allow us to determine the previously missing PPN parameter \(\sigma_-^{IJ}\) through the choice of a suitable gauge fixing. The relevant components of the field equations we need to consider are given by
\begin{subequations}\label{eqn:o4eom}
\begin{align}
K^{I(4)}_{00} &= 8\pi T^{I(4)}_{00}\,,\\
K^{I(4)}_{\alpha\beta} &= 8\pi T^{I(4)}_{\alpha\beta}\,.
\end{align}
\end{subequations}
As in the previous sections we start our calculation by expanding the components \(K^{I(4)}_{00}, K^{I(4)}_{\alpha\beta}\) of the curvature tensors to their most general form compatible with our assumptions~\textit{(\ref{ass:fields})}--\textit{(\ref{ass:flat})}. We find that these can be written as
\begin{subequations}\label{eqn:o4curvature}
\begin{align}
K^{I(4)}_{00} &= \sum_{J = 1}^{N}\left(S^{IJ}_{1}h^{J(4)}_{00,\alpha\alpha} + S^{IJ}_{2}h^{J(4)}_{\alpha\alpha,\beta\beta} + S^{IJ}_{3}h^{J(4)}_{\alpha\beta,\alpha\beta} + S^{IJ}_{4}h^{J(3)}_{0\alpha,0\alpha} + S^{IJ}_{5}h^{J(2)}_{00,00} + S^{IJ}_{6}h^{J(2)}_{\alpha\alpha,00}\right) + Q^I_{00}\,,\label{eqn:o4curvature1}\\
K^{I(4)}_{\alpha\beta} &= \sum_{J = 1}^{N}\left(T^{IJ}_{1}h^{J(4)}_{00,\alpha\beta} + T^{IJ}_{2}h^{J(4)}_{\gamma\gamma,\alpha\beta} + T^{IJ}_{3}h^{J(4)}_{\alpha\beta,\gamma\gamma} + T^{IJ}_{4}h^{J(4)}_{\gamma(\alpha,\beta)\gamma} + T^{IJ}_{5}h^{J(3)}_{0(\alpha,\beta)0} + T^{IJ}_{6}h^{J(2)}_{\alpha\beta,00}\right)\nonumber\\
&\phantom{=}+ \sum_{J = 1}^{N}\left(C^{IJ}_{1}h^{J(4)}_{00,\gamma\gamma} + C^{IJ}_{2}h^{J(4)}_{\gamma\gamma,\delta\delta} + C^{IJ}_{3}h^{J(4)}_{\gamma\delta,\gamma\delta} + C^{IJ}_{4}h^{J(3)}_{0\gamma,0\gamma} + C^{IJ}_{5}h^{J(2)}_{00,00} + C^{IJ}_{6}h^{J(2)}_{\gamma\gamma,00}\right)\delta_{\alpha\beta} + Q^I_{\alpha\beta}\label{eqn:o4curvature2}\,.
\end{align}
\end{subequations}
The constants \(S^{IJ}_{1}, \ldots, S^{IJ}_{6}, C^{IJ}_{1}, \ldots, C^{IJ}_{6}, T^{IJ}_{1}, \ldots, T^{IJ}_{6}\), some of which we already encountered in the expansion~\eqref{eqn:o2curvature} of the curvature tensors at the second velocity order, are uniquely determined by the concrete multimetric gravity theory under consideration, and can again be calculated from the linearized gravitational field equations. Again we see that each metric perturbation at the fourth velocity order depends on the lower order perturbations of all metrics. This also applies to the terms \(Q^I_{00}\) and \(Q^I_{\alpha\beta}\), which involve quadratic combinations of the second velocity order perturbations \(h^{J(2)}_{00}\) and \(h^{J(2)}_{\alpha\beta}\) of all metrics \(g^J_{ab}\). They can be calculated from the full, non-linear field equations of a concrete multimetric gravity theory. A detailed analysis shows that \(Q^I_{00}\) can be expanded into 18 different combinations of the second order metric perturbations, while \(Q^I_{\alpha\beta}\) similarly expands into 53 different terms. We omit these expansions here for brevity. Using the expression~\eqref{eqn:ppnmetric} for the metric perturbation we can write \(Q^I_{00}\) in the form
\begin{equation}\label{eqn:q00}
Q^I_{00} = \sum_{J,K = 1}^{N}\left(Z^{IJK}_{1}\triangle\chi^J\triangle\triangle\chi^K + Z^{IJK}_{2}\chi^J_{,\alpha\beta}\triangle\chi^K_{,\alpha\beta} + Z^{IJK}_{3}\triangle\chi^J_{,\alpha}\triangle\chi^K_{,\alpha} + Z^{IJK}_{4}\chi^J_{,\alpha\beta\gamma}\chi^K_{,\alpha\beta\gamma}\right)\,,
\end{equation}
where the constants \(Z^{IJK}_{1}, \ldots, Z^{IJK}_{4}\) can be calculated from the full field equations of a concrete multimetric gravity theory and involve the previously calculated PPN parameters \(\alpha^{IJ}, \gamma^{IJ}, \theta^{IJ}\). We further insert our solutions for the metric perturbation at the second and third velocity order and obtain
\begin{equation}\label{eqn:somega}
\sum_{J = 1}^{N}\left(S^{IJ}_{4}h^{J(3)}_{0\alpha,0\alpha} + S^{IJ}_{5}h^{J(2)}_{00,00} + S^{IJ}_{6}h^{J(2)}_{\alpha\alpha,00}\right) = \sum_{J = 1}^{N}H^{IJ}_{1}\triangle\left(\Omega_2^J + \Omega_3^J - \Omega_1^J\right)\,,
\end{equation}
where we used the properties of the PPN potentials and introduced another constant
\begin{equation}\label{eqn:coeffh1}
H^{IJ}_{1} = \sum_{K = 1}^{N}\left(S^{IK}_{4}\sigma_-^{KJ} - S^{IK}_{5}\alpha^{KJ} - S^{IK}_{6}(3\gamma^{KJ} + \theta^{KJ})\right)\,.
\end{equation}
Inserting the results~\eqref{eqn:q00} and~\eqref{eqn:somega} into the expansion~\eqref{eqn:o4curvature1} for \(K^{I(4)}_{00}\) we are left with the term \(h^{I(4)}_{00,\alpha\alpha}\), which is the one we would like to solve for, as well as the pure trace term \(h^{I(4)}_{\alpha\alpha,\beta\beta}\) and the double divergence \(h^{I(4)}_{\alpha\beta,\alpha\beta}\), which are also unknown, but not relevant for our post-Newtonian approximation~\eqref{eqn:ppnmetric}. In order to eliminate the latter two terms we consider the trace \(K^{I(4)}_{\alpha\alpha}\) and double divergence \(K^{I(4)}_{\alpha\beta,\alpha\beta}\) of the fourth order curvature components~\eqref{eqn:o4curvature2}, which take the form
\begin{subequations}
\begin{align}
K^{I(4)}_{\alpha\alpha} &= \sum_{J = 1}^{N}\left(\bar{C}^{IJ}_{1}\triangle h^{J(4)}_{00} + \bar{C}^{IJ}_{2}\triangle h^{J(4)}_{\alpha\alpha} + \bar{C}^{IJ}_{3}h^{J(4)}_{\alpha\beta,\alpha\beta} + H^{IJ}_{2}\triangle\left(\Omega_2^J + \Omega_3^J - \Omega_1^J\right)\right) + Q^I_{\alpha\alpha}\,,\\
K^{I(4)}_{\alpha\beta,\alpha\beta} &= \triangle\sum_{J = 1}^{N}\left(\bar{T}^{IJ}_{1}\triangle h^{J(4)}_{00} + \bar{T}^{IJ}_{2}\triangle h^{J(4)}_{\alpha\alpha} + \bar{T}^{IJ}_{3}h^{J(4)}_{\alpha\beta,\alpha\beta} + H^{IJ}_{3}\triangle\left(\Omega_2^J + \Omega_3^J - \Omega_1^J\right)\right) + Q^I_{\alpha\beta,\alpha\beta}\,.
\end{align}
\end{subequations}
The coefficients of the unknown fourth order metric components are given by the linear combinations
\begin{subequations}
\begin{gather}
\bar{C}^{IJ}_{1} = 3C^{IJ}_{1} + T^{IJ}_{1}\,, \quad
\bar{C}^{IJ}_{2} = 3C^{IJ}_{2} + T^{IJ}_{2} + T^{IJ}_{3}\,, \quad
\bar{C}^{IJ}_{3} = 3C^{IJ}_{3} + T^{IJ}_{4}\,,\\
\bar{T}^{IJ}_{1} = C^{IJ}_{1} + T^{IJ}_{1}\,, \quad
\bar{T}^{IJ}_{2} = C^{IJ}_{2} + T^{IJ}_{2}\,, \quad
\bar{T}^{IJ}_{3} = C^{IJ}_{3} + T^{IJ}_{3} + T^{IJ}_{4}\,.
\end{gather}
\end{subequations}
The terms involving the PPN potentials \(\Omega_1^I, \ldots, \Omega_3^I\) originate from inserting the solutions for the second and third order metric perturbations in analogy to equation~\eqref{eqn:somega}. Their coefficients are given by the constants
\begin{subequations}\label{eqn:coeffh23}
\begin{align}
H^{IJ}_{2} &= \sum_{K = 1}^{N}\left((3C^{IK}_{4} + T^{IK}_{5})\sigma_-^{KJ} - 3C^{IK}_{5}\alpha^{KJ} - (3C^{IK}_{6} + T^{IK}_{6})(3\gamma^{KJ} + \theta^{KJ})\right)\,,\\
H^{IJ}_{3} &= \sum_{K = 1}^{N}\left((C^{IK}_{4} + T^{IK}_{5})\sigma_-^{KJ} - C^{IK}_{5}\alpha^{KJ} - C^{IK}_{6}(3\gamma^{KJ} + \theta^{KJ}) - T^{IK}_{6}(\gamma^{KJ} - \theta^{KJ})\right)\,.
\end{align}
\end{subequations}
The quadratic terms \(Q^I_{\alpha\alpha}\) and \(Q^I_{\alpha\beta,\alpha\beta}\) can now be expanded in analogy to equation~\eqref{eqn:q00}. Their expansions take the form
\begin{subequations}\label{eqn:qab}
\begin{align}
Q^I_{\alpha\alpha} &= \sum_{J,K = 1}^{N}\left(D^{IJK}_{1}\triangle\chi^J\triangle\triangle\chi^K + D^{IJK}_{2}\chi^J_{,\alpha\beta}\triangle\chi^K_{,\alpha\beta} + D^{IJK}_{3}\triangle\chi^J_{,\alpha}\triangle\chi^K_{,\alpha} + D^{IJK}_{4}\chi^J_{,\alpha\beta\gamma}\chi^K_{,\alpha\beta\gamma}\right)\,,\label{eqn:qaa}\\
Q^I_{\alpha\beta,\alpha\beta} &= \sum_{J,K = 1}^{N}\big(E^{IJK}_{1}\triangle\chi^J\triangle\triangle\triangle\chi^K + E^{IJK}_{2}\chi^J_{,\alpha\beta}\triangle\triangle\chi^K_{,\alpha\beta} + E^{IJK}_{3}\triangle\chi^J_{,\alpha}\triangle\triangle\chi^K_{,\alpha}\label{eqn:qabab}\\
&\phantom{=}+ E^{IJK}_{4}\chi^J_{,\alpha\beta\gamma}\triangle\chi^K_{,\alpha\beta\gamma} + E^{IJK}_{5}\triangle\triangle\chi^J\triangle\triangle\chi^K + E^{IJK}_{6}\triangle\chi^J_{,\alpha\beta}\triangle\chi^K_{,\alpha\beta} + E^{IJK}_{7}\chi^J_{,\alpha\beta\gamma\delta}\chi^K_{,\alpha\beta\gamma\delta}\big)\nonumber\,,
\end{align}
\end{subequations}
with constants \(D^{IJK}_{1}, \ldots, D^{IJK}_{4}, E^{IJK}_{1}, \ldots, E^{IJK}_{7}\) that can be calculated from the full field equations of a concrete multimetric gravity theory and involve the PPN parameters \(\alpha^{IJ}, \gamma^{IJ}, \theta^{IJ}\) in analogy to the constants \(Z^{IJK}_{1}, \ldots, Z^{IJK}_{4}\) in equation~\eqref{eqn:q00}. The quadratic terms that appear in these expansions can be expressed in terms of the non-linear PPN potentials \(\Psi_A^{IJK}\). For the terms in equations~\eqref{eqn:q00} and~\eqref{eqn:qaa} we find
\begin{gather}
\triangle\chi^I\triangle\triangle\chi^J = \triangle\left(\Psi_1^{IJ} + 2\Psi_3^{IJ} + \Psi_5^{IJ}\right)\,, \quad
\chi^I_{,\alpha\beta}\triangle\chi^J_{,\alpha\beta} = \triangle\left(\Psi_2^{IJ} + 2\Psi_4^{IJ} + \Psi_6^{IJ}\right)\,,\nonumber\\
\triangle\chi^I_{,\alpha}\triangle\chi^J_{,\alpha} = \triangle\left(\Psi_3^{IJ} + \Psi_3^{JI} + 2\Psi_6^{IJ}\right)\,, \quad
\chi^I_{,\alpha\beta\gamma}\chi^J_{,\alpha\beta\gamma} = \triangle\left(\Psi_4^{IJ} + \Psi_4^{JI} + 2\Psi_7^{IJ}\right)\,,
\end{gather}
while the terms in equation~\eqref{eqn:qabab} can directly be read off from the definition~\eqref{eqn:psipot} of the non-linear PPN potentials. We need to perform a similar expansion of the matter side of the field equations~\eqref{eqn:o4eom} in terms of the PPN potentials. The relevant components of the energy-momentum tensors~\eqref{eqn:energymomentum} are given by
\begin{subequations}
\begin{align}
T^{I(4)}_{00} &= \rho^I\left(\Pi^I + {v^I}^2 + \sum_{J = 1}^{N}\alpha^{IJ}\triangle\chi^J\right)\,,\\
T^{I(4)}_{\alpha\alpha} &= \rho^I{v^I}^2 + 3p^I\,,\\
T^{I(4)}_{\alpha\beta,\alpha\beta} &= (\rho^Iv^I_{\alpha}v^I_{\beta})_{,\alpha\beta} + \triangle p^I\,.
\end{align}
\end{subequations}
A comparison with the definitions of the PPN potentials in section~\ref{subsec:ppnmetric} yields the relations
\begin{equation}
\triangle\Phi_{\Pi}^I = -4\pi\rho^I\Pi^I\,, \quad
\triangle\Phi_p^I = -4\pi p^I\,, \quad
\triangle\Omega_1^I = -4\pi\rho^I{v^I}^2\,, \quad
\triangle\triangle\Omega_2^I = -8\pi(\rho^Iv^I_{\alpha}v^I_{\beta})_{,\alpha\beta}\,.
\end{equation}
We can now eliminate the unknown and irrelevant terms \(h^{I(4)}_{\alpha\alpha}\) and \(h^{I(4)}_{\alpha\beta,\alpha\beta}\) by choosing a suitable linear combination of the curvature terms \(K^{I(4)}_{00}, K^{I(4)}_{\alpha\alpha}, K^{I(4)}_{\alpha\beta,\alpha\beta}\) to obtain
\begin{multline}\label{eqn:finalo4}
\sum_{J = 1}^{N}\left(M_{1}^{IJ}\triangle K^{J(4)}_{00} + M_{2}^{IJ}\triangle K^{J(4)}_{\alpha\alpha} + M_{3}^{IJ}K^{J(4)}_{\alpha\beta,\alpha\beta}\right) =\\
8\pi\sum_{J = 1}^{N}\left(M_{1}^{IJ}\triangle T^{J(4)}_{00} + M_{2}^{IJ}\triangle T^{J(4)}_{\alpha\alpha} + M_{3}^{IJ}T^{J(4)}_{\alpha\beta,\alpha\beta}\right)\,,
\end{multline}
where the constants \(M^{IJ}_{1}, \ldots, M^{IJ}_{3}\) are chosen so that the coefficient of the term \(h^{I(4)}_{00}\) becomes \(\delta^{IJ}\) and the spatial components \(h^{I(4)}_{\alpha\beta}\) of the fourth order perturbations cancel. Comparison of the coefficients of the PPN potentials in the metric~\eqref{eqn:ppnmetric} and equation~\eqref{eqn:finalo4} we can then read off the PPN parameters
\begin{subequations}
\begin{align}
\phi_p^{IJ} &= -6M^{IJ}_{2} - 2M^{IJ}_{3} + 2\sum_{L = 1}^{N}\left[M^{IL}_{1}H^{LJ}_1 + M^{IL}_{2}H^{LJ}_2 + M^{IL}_{3}H^{LJ}_3\right]\,,\\
\phi_{\Pi}^{IJ} &= -2M^{IJ}_{1}\,,\\
\omega_1^{IJ} &= -2M^{IJ}_{1} - 2M^{IJ}_{2} + \sum_{L = 1}^{N}\left[M^{IL}_{1}H^{LJ}_1 + M^{IL}_{2}H^{LJ}_2 + M^{IL}_{3}H^{LJ}_3\right]\,,\\
\omega_2^{IJ} &= -M^{IJ}_{3} - \sum_{L = 1}^{N}\left[M^{IL}_{1}H^{LJ}_1 + M^{IL}_{2}H^{LJ}_2 + M^{IL}_{3}H^{LJ}_3\right]\,,\\
\psi_1^{IJK} &= M^{IK}_{1}\alpha^{KJ} - \sum_{L = 1}^{N}\left[M^{IL}_{1}Z^{LJK}_{1} + M^{IL}_{2}D^{LJK}_{1} + M^{IL}_{3}E^{LJK}_{1}\right]\,,\\
\psi_2^{IJK} &= -\sum_{L = 1}^{N}\left[M^{IL}_{1}Z^{LJK}_{2} + M^{IL}_{2}D^{LJK}_{2} + M^{IL}_{3}E^{LJK}_{2}\right]\,,\\
\psi_3^{IJK} &= 2M^{IK}_{1}\alpha^{KJ} - \frac{1}{2}\sum_{L = 1}^{N}\left[M^{IL}_{1}H^{LK}_1 + M^{IL}_{2}H^{LK}_2 + M^{IL}_{3}H^{LK}_3\right]\alpha^{KJ}\\
&\phantom{=}- \sum_{L = 1}^{N}\left[M^{IL}_{1}(2Z^{LJK}_{1} + Z^{LJK}_{3} + Z^{LKJ}_{3}) + M^{IL}_{2}(2D^{LJK}_{1} + D^{LJK}_{3} + D^{LKJ}_{3}) + M^{IL}_{3}E^{LJK}_{3}\right]\,,\nonumber\\
\psi_4^{IJK} &= -\sum_{L = 1}^{N}\left[M^{IL}_{1}(2Z^{LJK}_{2} + Z^{LJK}_{4} + Z^{LKJ}_{4}) + M^{IL}_{2}(2D^{LJK}_{2} + D^{LJK}_{4} + D^{LKJ}_{4}) + M^{IL}_{3}E^{LJK}_{4}\right]\,,\\
\psi_5^{IJK} &= M^{IK}_{1}\alpha^{KJ} - \frac{1}{2}\sum_{L = 1}^{N}\left[M^{IL}_{1}H^{LJ}_1 + M^{IL}_{2}H^{LJ}_2 + M^{IL}_{3}H^{LJ}_3\right]\alpha^{JK}\\
&\phantom{=}- \sum_{L = 1}^{N}\left[M^{IL}_{1}Z^{LJK}_{1} + M^{IL}_{2}D^{LJK}_{1} + M^{IL}_{3}E^{LJK}_{5}\right]\,,\nonumber\\
\psi_6^{IJK} &= -\sum_{L = 1}^{N}\left[M^{IL}_{1}(Z^{LJK}_{2} + 2Z^{LJK}_{3}) + M^{IL}_{2}(D^{LJK}_{2} + 2D^{LJK}_{3}) + M^{IL}_{3}E^{LJK}_{6}\right]\,,\\
\psi_7^{IJK} &= -\sum_{L = 1}^{N}\left[2M^{IL}_{1}Z^{LJK}_{4} + 2M^{IL}_{2}D^{LJK}_{4} + M^{IL}_{3}E^{LJK}_{7}\right]\,.
\end{align}
\end{subequations}
Note that they still depend on the so far undetermined PPN parameter \(\sigma_-^{IJ}\), which enters these equations through the constants \(H_1^{IJ}, \ldots, H_3^{IJ}\). Using the gauge condition \(\psi_1^{III} = \psi_5^{III}\) introduced in section~\ref{subsec:ppngauge} we can finally solve for all PPN parameters.

This completes our calculation of the PPN parameters for a general multimetric gravity theory. The result we obtained depends on a number of constant coefficients that appear in the gravitational field equations and characterize the theory under consideration. We will explicitly calculate these coefficients for two example theories and derive their PPN parameters in the following section.

\section{Applications}\label{sec:applications}
In the preceding section we have presented a procedure for solving the post-Newtonian field equations of an arbitrary multimetric gravity theory satisfying assumptions~\textit{(\ref{ass:fields})}--\textit{(\ref{ass:flat})} listed in the introduction. We now apply this procedure to two gravity theories and calculate their post-Newtonian limits. The first theory displayed in section~\ref{subsec:genrel} is general relativity, which is in fact a theory of only one metric, but will serve as an illustrative example. We show that our procedure reproduces its well-known PPN parameters. We then discuss a wide class of multimetric gravity theories with an arbitrary number \(N\) of metric tensors in section~\ref{subsec:repgrav}. From a comparison of the calculated PPN parameters with their experimentally measured values we obtain conditions on this class of theories. We study which theories satisfy these conditions and are thus compatible with post-Newtonian experiments in gravitational physics. Finally, we discuss the implications of our results for cosmology.

\subsection{General relativity}\label{subsec:genrel}
The first and rather illustrative example we discuss in this section is general relativity, which is defined by the Einstein-Hilbert action
\begin{equation}
S_G = \frac{1}{16\pi}\int d^4x\,\sqrt{g}R\,.
\end{equation}
Since there is only a single metric \(g_{ab}\) we will drop all indices \(I, J, \ldots\) in this section. The field equations are the familiar Einstein equations
\begin{equation}\label{eqn:ehaction}
K_{ab} = R_{ab} - \frac{1}{2}Rg_{ab} = 8\pi T_{ab}\,,
\end{equation}
where the curvature tensor \(K_{ab}\) is simply given by the Einstein tensor. In order to solve the field equations up to the post-Newtonian level and calculate the metric perturbation~\eqref{eqn:ppnmetric} we follow the steps detailed in section~\ref{sec:calculation}. First we calculate the components \(K^{(2)}_{00}\) and \(K^{(2)}_{\alpha\beta}\) of the curvature tensor at the second velocity order \(\mathcal{O}(2)\). By comparison with the most general second order curvature tensor~\eqref{eqn:o2curvature} we can read off the coefficients
\begin{equation}\label{eqn:gr2coeff}
S_1 = 0\,, \quad -S_2 = S_3 = -C_1 = C_2 = -C_3 = T_1 = -T_2 = -T_3 = \frac{1}{2}\,, \quad T_4 = 1\,.
\end{equation}
Inserting these into equation~\eqref{eqn:coeffx123} we find the constants
\begin{equation}
X_1 = \gamma + \theta\,, \quad X_2 = -X_3 = \frac{\alpha - \gamma - \theta}{2}\,.
\end{equation}
In order to determine the PPN parameters \(\alpha, \gamma, \theta\) we need to solve the linear equations
\begin{equation}
X_1 = 1\,, \quad X_2 = X_3 = 0\,.
\end{equation}
As explained in section~\ref{subsec:ppno2} they are linearly dependent as a consequence of the diffeomorphism invariance of the gravitational action~\eqref{eqn:ehaction}. We therefore need to fix a gauge as shown in section~\ref{subsec:ppngauge} using the condition \(\theta = 0\). With this gauge fixing we find the PPN parameters \(\alpha = 1\) and \(\gamma = 1\).

In the next step we solve the gravitational field equations up to the third velocity order. For this purpose we need to calculate the component \(K^{(3)}_{0\alpha}\) of the curvature tensor. Comparison with its most general form~\eqref{eqn:o3curvature} yields the coefficients
\begin{equation}
-V_1 = V_2 = -V_4 = V_5 = \frac{1}{2}\,, \quad V_3 = 0\,.
\end{equation}
We insert these values into equation~\eqref{eqn:coeffx45} and obtain the constants
\begin{equation}
X_4 = -\frac{\sigma_+}{2}\,, \quad X_5 = \gamma + \theta\,.
\end{equation}
Again we find that the resulting equations
\begin{equation}
X_4 = X_5 = 1
\end{equation}
are linearly dependent on the previously solved equations at second velocity order as another consequence of the diffeomorphism invariance of the underlying gravity theory. In this case we find that the equation for \(X_5\) is identically solved by the second order solution for \(\gamma\) and \(\theta\), while the equation for \(X_4\) contains only the parameter \(\sigma_+\). We therefore only obtain its value \(\sigma_+ = -2\) and need to defer the calculation of \(\sigma_-\) to the remaining fourth order calculation.

For the final part of our procedure we need to calculate the curvature tensor up to the fourth velocity order \(\mathcal{O}(4)\). Writing the components \(K^{(4)}_{00}\) and \(K^{(4)}_{\alpha\beta}\) in the form~\eqref{eqn:o4curvature} we can read off the coefficients
\begin{equation}
S_4 = S_5 = S_6 = C_5 = 0\,, \quad C_4 = -T_5 = 1\,, \quad -C_6 = T_6 = \frac{1}{2}\,,
\end{equation}
in addition to the coefficients~\eqref{eqn:gr2coeff} that we already determined from the second velocity order calculation. Inserting these into equations~\eqref{eqn:coeffh1} and~\eqref{eqn:coeffh23} we find the constants
\begin{equation}
H_1 = 0\,, \quad H_2 = 3\gamma + \theta + 2\sigma_-\,, \quad H_3 = \gamma + \theta\,.
\end{equation}
We further need to determine the quadratic terms \(Q_{00}\) and \(Q_{\alpha\beta}\) from an expansion of the curvature tensor \(K_{ab}\) up to the quadratic order in the metric perturbation \(h_{ab}\). After inserting the post-Newtonian metric~\eqref{eqn:ppnmetric} we can compare the result with equations~\eqref{eqn:q00} and~\eqref{eqn:qab}, from which we obtain the coefficients
\begin{gather}
Z_1 = (\alpha + 2\gamma + \theta)(\gamma + \theta)\,, \quad Z_2 = -\theta(\gamma + \theta)\,, \quad Z_3 = \frac{3\gamma^2 + 2\gamma\theta - 3\theta^2}{4}\,, \quad Z_4 = \frac{\theta^2}{2}\,,\nonumber\\
D_3 = \frac{3(\theta^2 - \gamma^2) - 2\gamma\theta - 2\alpha(\alpha + \gamma + 3\theta)}{4}\,, \quad E_5 = \frac{\alpha(2\theta - 2\gamma - \alpha) + (\gamma + \theta)(\gamma - 3\theta)}{4}\,,\\
E_3 = \frac{\alpha(\theta - 2\gamma) + (\gamma + \theta)(\gamma - 2\theta)}{2}\,, \quad E_4 = \frac{3\theta(\gamma + \theta - \alpha)}{2}\,, \quad E_6 = \frac{(\alpha - \gamma - \theta)(\alpha - \gamma - 5\theta)}{4}\,,\nonumber\\
D_1 = \alpha(\theta - \alpha) - (\gamma + \theta)^2\,, \quad D_2 = \theta(2\gamma + 2\theta - 3\alpha)\,, \quad D_4 = -\frac{\theta^2}{2}\,, \quad E_1 = E_2 = E_7 = 0\,.\nonumber
\end{gather}
Inserting the previously found solution for the PPN parameters \(\alpha, \gamma, \theta\) yields
\begin{gather}
H_2 = 3 + 2\sigma_-\,, \quad H_3 = 1\,, \quad Z_1 = 3\,, \quad Z_3 = \frac{3}{4}\,, \quad D_1 = -2\,, \quad D_3 = -\frac{7}{4}\,, \quad E_3 = E_5 = -\frac{1}{2}\,,\nonumber\\
H_1 = Z_2 = Z_4 = D_2 = D_4 = E_1 = E_2 = E_4 = E_6 = E_7 = 0\,.
\end{gather}
In order to solve the fourth order equations of motion for \(h^{(4)}_{00}\) we determine a suitable linear combination in the form~\eqref{eqn:finalo4}. Using the coefficients
\begin{equation}
M_1 = M_2 = -1\,, \quad M_3 = 0\,,
\end{equation}
we can read off the equations for the remaining PPN parameters
\begin{gather}
\omega_1 = 1 - 2\sigma_-\,, \quad \omega_2 = 3 + 2\sigma_-\,, \quad \phi_p = -4\sigma_-\,, \quad \phi_{\Pi} = 2\,, \quad \psi_3 = -\frac{1}{2} + \sigma_-\,,\nonumber\\
\psi_5 = \frac{3}{2} + \sigma_-\,, \quad \psi_6 = -2\,, \quad \psi_1 = \psi_2 = \psi_4 = \psi_7 = 0\,.
\end{gather}
In order to determine the PPN parameter \(\sigma_-\) and solve these equations we use the gauge condition \(\psi_1 = \psi_5\) introduced in section~\ref{subsec:ppngauge}. This finally yields us the complete set of PPN parameters
\begin{gather}
\alpha = \gamma = 1\,, \quad \sigma_+ = \psi_3 = \psi_6 = -2\,, \quad \sigma_- = -\frac{3}{2}\,, \quad \phi_{\Pi} = 2\,, \quad \phi_p = 6\,,\nonumber\\
\omega_1 = 4\,, \quad \theta = \omega_2 = \psi_1 = \psi_2 = \psi_4 = \psi_5 = \psi_7 = 0\,.
\end{gather}
By comparison with the measured values~\eqref{eqn:measured} of the PPN parameters for visible matter we see that they agree with the values we calculated in this section. Our formalism thus reproduces the well-known result that general relativity is consistent with solar system experiments at the post-Newtonian level.

\subsection{Multimetric repulsive gravity}\label{subsec:repgrav}
The second example we study here is a class of multimetric gravity theories with \(N \geq 2\) metric tensors and a corresponding number of standard model copies. It is defined by the action
\begin{align}
S_G &= \frac{1}{16\pi}\int d^4x\,\sqrt{g_0}\Bigg[\sum_{I = 1}^{N}\Big(c_{1}R^I + g^{I\,ij}\left(c_{3}\tilde{S}^I{}_i\tilde{S}^I{}_j + c_{5}\tilde{S}^I{}_k\tilde{S}^{I\,k}{}_{ij} + c_{7}\tilde{S}^{I\,k}{}_{il}\tilde{S}^{I\,l}{}_{jk}\right)\nonumber\\
&\phantom{=}+ g^{I\,ij}g^{I\,kl}g^I{}_{mn}\left(c_{9}\tilde{S}^{I\,m}{}_{ik}\tilde{S}^{I\,n}{}_{jl} + c_{11}\tilde{S}^{I\,m}{}_{ij}\tilde{S}^{I\,n}{}_{kl}\right)\Big)\nonumber\\
&\phantom{=}+ \sum_{I,J = 1}^{N}\Big(c_{2}g^{I\,ij}R^J{}_{ij} + g^{I\,ij}\left(c_{4}S^{IJ}{}_iS^{IJ}{}_j + c_{6}S^{IJ}{}_kS^{IJ\,k}{}_{ij} + c_{8}S^{IJ\,k}{}_{il}S^{IJ\,l}{}_{jk}\right)\label{eqn:repaction}\\
&\phantom{=}+ g^{I\,ij}g^{I\,kl}g^I{}_{mn}\left(c_{10}S^{IJ\,m}{}_{ik}S^{IJ\,n}{}_{jl} + c_{12}S^{IJ\,m}{}_{ij}S^{IJ\,n}{}_{kl}\right)\Big)\Bigg]\,,\nonumber
\end{align}
where \(c_{1}, \ldots, c_{12}\) are constant parameters and we used the connection difference tensors
\begin{equation}
S^{IJ\,i}{}_{jk} = \Gamma^{I\,i}{}_{jk} - \Gamma^{J\,i}{}_{jk}\,, \quad
S^{IJ}{}_j = S^{IJ\,k}{}_{jk}\,, \quad
\tilde{S}^{J\,i}{}_{jk} = \frac{1}{N}\sum_{I = 1}^{N}S^{IJ\,i}{}_{jk}\,,\quad
\tilde{S}^{J}{}_j = \tilde{S}^{J\,k}{}_{jk}
\end{equation}
and the mixed density
\begin{equation}
g_0 = \prod_{I = 1}^{N}\left(g^I\right)^{\frac{1}{N}}\,.
\end{equation}
Special cases of this action have been studied in the contexts of cosmology~\cite{Hohmann:2010vt} and gravitational waves~\cite{Hohmann:2011gb}, and a subset of their PPN parameters has been calculated~\cite{Hohmann:2010ni}. Using the formalism presented in this article we can now generalize these previous results to the action~\eqref{eqn:repaction} and calculate the full set of PPN parameters. For brevity we will only sketch this calculation. We will further restrict ourselves to theories which are consistent with the measured values~\eqref{eqn:measured} of the PPN parameters in the solar system. Instead of displaying the full result for the PPN parameters we will therefore discuss which restrictions we obtain on the parameters \(c_{1}, \ldots, c_{12}\) and display the PPN parameters for this restricted case.

An important property of the action~\eqref{eqn:repaction} is its symmetry with respect to arbitrary permutations of the sectors \((g^I,\varphi^I)\). This can be understood as a generalized Copernican principle, in the sense that the equations of motion both for gravity and matter are the same in each of the \(N\) sectors. As a consequence of this symmetry all constants \(P^{I_1 \cdots I_n}\) characterizing the theory, and thus in particular the PPN parameters and the expansion coefficients used in section~\ref{sec:calculation}, obey the same permutation symmetry, i.e., they satisfy the relation
\begin{equation}\label{eqn:symmetry}
P^{I_1 \cdots I_n} = \sum_{J_1, \ldots, J_n = 1}^{N}P^{J_1 \cdots J_n}\pi^{I_1J_1}\ldots\pi^{I_nJ_n}
\end{equation}
for arbitrary permutation matrices \(\pi^{IJ}\). Constants \(P_2^{IJ}\) and \(P_3^{IJK}\) with \(2\) and \(3\) indices hence take the most general form
\begin{equation}\label{eqn:symnotation}
P_2^{IJ} = \frac{\bar{P}_2}{N} + \hat{P}_2\delta^{IJ}\,, \quad P_3^{IJK} = \frac{\bar{P}_3}{N^2} + \frac{\ola{P}_3\delta^{IJ} + \ora{P}_3\delta^{IK} + \tilde{P}_3\delta^{JK}}{N} + \hat{P}_3\delta^{IJ}\delta^{IK}\,,
\end{equation}
and can thus be expressed by the tuples \((\bar{P}_2,\hat{P}_2)\) or \((\bar{P}_3,\ola{P}_3,\ora{P}_3,\tilde{P}_3,\hat{P}_3)\), respectively. We will make use of this notation for the remainder of this section.

For the Newtonian limit given by the metric perturbation~\eqref{eqn:h200} this symmetry implies that all diagonal elements \(\alpha^{II}\) are equal, which means that the effective gravitational constant is the same in all sectors \((g^I, \varphi^I)\). They can thus be rescaled to \(\alpha^{II} = 1\), in analogy to the parameter \(\alpha\) in the standard PPN metric~\eqref{eqn:standardppn}. It further follows that also all off-diagonal elements \(\alpha^{IJ}\) for \(I \neq J\), which determine the Newtonian gravitational force between the different matter types \(\varphi^I\), must be equal. Denoting their common value by \(\alpha^{IJ} = z\) and using the notation introduced in equation~\eqref{eqn:symnotation} we thus find
\begin{equation}\label{eqn:multinewton}
\bar{\alpha} = Nz\,, \quad \hat{\alpha} = 1 - z\,.
\end{equation}
We are particularly interested in gravity theories in which the different standard model copies \(\varphi^I\) mutually repel each other in the Newtonian limit, and in which this repulsive gravitational force is of equal strength compared to the attractive gravitational force within each matter sector. This corresponds to the case \(z = -1\). In the following derivation we will keep \(z\) arbitrary and discuss the particular consequences of the value \(z = -1\) later towards the end of this section.

Similar symmetry considerations as for \(\alpha^{IJ}\) also apply to all other PPN parameters that appear in the metric perturbations~\eqref{eqn:ppnmetric}. They are constants carrying \(2\) or \(3\) indices and can hence be written in the form~\eqref{eqn:symnotation}. Thus, the diagonal elements are equal for each of the PPN parameters. By comparison with the measured values~\eqref{eqn:standardparams} in the visible sector \(I = 1\) we then derive the experimental consistency conditions
\begin{equation}
\frac{\bar{\gamma}}{N} + \hat{\gamma} = 1\,, \quad \frac{\bar{\psi}_1}{N^2} + \frac{\ola{\psi}_1 + \ora{\psi}_1 + \tilde{\psi}_1}{N} + \hat{\psi}_1 = 0\,,
\end{equation}
and analogue expressions for the remaining PPN parameters. Similarly, the gauge conditions discussed in section~\ref{subsec:ppngauge} take the form
\begin{equation}
\frac{\bar{\theta}}{N} + \hat{\theta} = 0\,, \quad \frac{\bar{\psi}_1}{N^2} + \frac{\ola{\psi}_1 + \ora{\psi}_1 + \tilde{\psi}_1}{N} + \hat{\psi}_1 = \frac{\bar{\psi}_5}{N^2} + \frac{\ola{\psi}_5 + \ora{\psi}_5 + \tilde{\psi}_5}{N} + \hat{\psi}_5\,.
\end{equation}
Using these conditions we can calculate the PPN parameters for the multimetric gravity theory defined by the action~\eqref{eqn:repaction}. Consistency with the experimentally measured values~\eqref{eqn:standardparams} and the Newtonian limit~\eqref{eqn:multinewton} then yields the conditions
\begin{subequations}\label{eqn:cond}
\begin{align}
c_{1} &= \frac{(2 - N)(z - 1)}{4(z + N - 1)}(2c_{4} - c_{6} + c_{8}) + \frac{13N + 6 + (19N - 6)z}{4(z + N - 1)}c_{10}\nonumber\\
&\phantom{=}+ \frac{1 - 3N + (N^2 - 2N - 2)z - (2N^2 - 5N - 1)z^2}{4(z - 1)(z + N - 1)(Nz - z + 1)}\,,\\
c_{2} &= \frac{(N - 2)(z - 1)}{4(z + N - 1)}(2c_{4} - c_{6} + c_{8}) - \frac{13N + 6 + (19N - 6)z}{4(z + N - 1)}c_{10} + \frac{3 - N + (2N - 3)z}{4(z - 1)(z + N - 1)}\,,\\
c_{3} &= \frac{3N^2 - 19N + 14 - (13N^2 - 28N + 28)z - (2N^2 + 9N - 14)z^2}{6(z + N - 1)(Nz - z + 1)}c_{4}\nonumber\\
&\phantom{=}+ \frac{(N - 2)(z - 1)(3N - 1 + (2N + 1)z)}{12(z + N - 1)(Nz - z + 1)}(c_{6} - c_{8})\nonumber\\
&\phantom{=}+ \frac{19N^2 + N + 18 + (39N^2 + 36N - 36)z + (38N^2 - 37N + 18)z^2}{12(z + N - 1)(Nz - z + 1)}c_{10}\nonumber\\
&\phantom{=}- \frac{N^2 + 18N - 15 + (4N^2 - 14N + 30)z + (4N^2 - 4N - 15)z^2}{12(z - 1)(z + N - 1)(Nz - z + 1)}\,,\\
c_{5} &= \frac{(2 - N)(z - 1)(6N - 5 + (N + 5)z)}{12(z + N - 1)(Nz - z + 1)}(2c_{4} + c_{8})\nonumber\\
&\phantom{=}- \frac{6N^2 + 7N - 14 + (19N^2 - 28N + 28)z - (N^2 - 21N + 14)z^2}{12(z + N - 1)(Nz - z + 1)}c_{6}\nonumber\\
&\phantom{=}+ \frac{38N^2 - 37N + 18 + (39N^2 + 36N - 36)z + (19N^2 + N + 18)z^2}{12(z + N - 1)(Nz - z + 1)}c_{10}\nonumber\\
&\phantom{=}- \frac{2N^2 + 3N - 3 - (N^2 - 2N - 6)z + (2N^2 - 5N - 3)z^2}{12(z - 1)(z + N - 1)(Nz - z + 1)}\,,\\
c_{7} &= \frac{(N - 2)(z - 1)(3N - 2 + (N + 2)z)}{4(z + N - 1)(Nz - z + 1)}(2c_{4} - c_{6})\nonumber\\
&\phantom{=}- \frac{3N^2 - 4 + (6N^2 - 8N + 8)z - (N^2 - 8N + 4)z^2}{4(z + N - 1)(Nz - z + 1)}c_{8}\nonumber\\
&\phantom{=}- \frac{19N^2 - 12N + 12 + (26N^2 + 24N - 24)z + (19N^2 - 12N + 12)z^2}{4(z + N - 1)(Nz - z + 1)}c_{10}\nonumber\\
&\phantom{=}+ \frac{N^2 + 7N - 6 + (N^2 - 4N + 12)z + (2N^2 - 3N - 6)z^2}{4(z - 1)(z + N - 1)(Nz - z + 1)}\,,\\
c_{9} &= \frac{(N - 2)(z - 1)(3N - 4 - (N - 4)z)}{12(z + N - 1)(Nz - z + 1)}(2c_{4} - c_{6} + c_{8})\nonumber\\
&\phantom{=}- \frac{19N^2 - 14N - 24 + (24N^2 - 48N + 48)z - (19N^2 - 62N + 24)z^2}{12(z + N - 1)(Nz - z + 1)}c_{10}\nonumber\\
&\phantom{=}+ \frac{N^2 + 9N - 12 + (7N^2 - 20N + 24)z - (2N^2 - 11N + 12)z^2}{12(z - 1)(z + N - 1)(Nz - z + 1)}
\end{align}
\end{subequations}
on the parameters \(c_{1}, \ldots, c_{12}\) in the multimetric gravity action~\eqref{eqn:repaction}. The full set of PPN parameters for this experimentally consistent gravity theory is displayed in appendix~\ref{sec:repparam}. Note that these parameters satisfy the conditions~\eqref{eqn:lorentzinv}, so that there are no preferred-frame effects. This is a consequence of our assumption~\textit{(\ref{ass:action})} that the vacuum solution is given by a set of flat metrics, which do not single out a preferred frame.

We now turn our focus from the physics of the solar system, where the dynamics of gravity is dominated by only a single standard model copy constituting the sun and the planets, to the cosmological dynamics of our model. The crucial assumption we make in this context is that on cosmological scales, much larger than the size of structures such as galaxies, the matter content of the universe can be modeled by a homogeneous and isotropic fluid, which is constituted by equal amounts of all standard model copies. The latter can be understood as a version of the Copernican principle, which states that no matter sector is distinguished, and which is also reflected by the symmetry of the action~\eqref{eqn:repaction} under permutations of the sectors. From this assumption follows that the gravitational interaction between the different matter sectors significantly influences the cosmological dynamics. In the following we study these dynamics and discuss in particular the case \(z = -1\) in which different standard model copies repel each other.

The assumption of large-scale homogeneity and isotropy implies that on cosmological scales the metrics must be of Robertson--Walker type,
\begin{equation}\label{eqn:flrw}
g^I = -(n^I)^2(t)dt \otimes dt + (a^I)^2(t)\gamma_{\alpha\beta}dx^{\alpha} \otimes dx^{\beta}\,,
\end{equation}
with lapse functions \(n^I(t)\), scale factors \(a^I(t)\), and a common purely spatial metric \(\gamma_{\alpha\beta}\) of constant curvature \(k \in \{-1, 0, 1\}\) and Riemann tensor \(R(\gamma)_{\alpha\beta\gamma\delta} = 2k\gamma_{\alpha[\gamma}\gamma_{\delta]\beta}\). Moreover, it follows that all matter is co-moving in the universe rest frame, \(\vec{v}^I = 0\). We further assume that we can neglect the internal energy \(\Pi^I = 0\) and apply the Copernican principle in the sense that all matter densities and pressures become equal, \(\rho^I = \rho\) and \(p^I = p\). From the symmetry of the action~\eqref{eqn:repaction} it then follows that the metrics \(g^I = g\), and thus in particular the scale factors \(a^I(t) = a(t)\) and lapse functions \(n^I(t) = n(t)\) become equal. By rescaling the cosmological time \(t\) we can then set the common lapse function to \(n(t) \equiv 1\). For this simple cosmological model the connection difference tensors \(S^{IJ\,i}{}_{jk}\) vanish and the gravitational field equations derived from the action~\eqref{eqn:repaction} reduce to
\begin{equation}
(c_1 + c_2)\left(R_{ab} - \frac{1}{2}Rg_{ab}\right) = 8\pi T_{ab}\,.
\end{equation}
Using the Robertson--Walker form~\eqref{eqn:flrw} of the metric we then find the cosmological equations of motion
\begin{subequations}\label{eqn:coseom}
\begin{align}
8\pi\rho &= 3(c_1 + c_2)\left(\frac{\dot{a}^2}{a^2} + \frac{k}{a^2}\right)\,,\label{eqn:density}\\
8\pi p &= -(c_1 + c_2)\left(2\frac{\ddot{a}}{a} + \frac{\dot{a}^2}{a^2} + \frac{k}{a^2}\right)\,,\label{eqn:pressure}
\end{align}
\end{subequations}
from which we derive the acceleration equation
\begin{equation}\label{eqn:accel}
\frac{\ddot{a}}{a} = -\frac{4\pi}{3(c_1 + c_2)}(\rho + 3p)\,.
\end{equation}
Using the conditions~\eqref{eqn:cond} on the parameters of our multimetric gravity model we find the simple expression
\begin{equation}
c_1 + c_2 = \frac{1}{1 + (N - 1)z}\,,
\end{equation}
which in the repulsive gravity case \(z = -1\) becomes singular for \(N = 2\) and negative for \(N \geq 3\). This reproduces our previous results that repulsive gravity cannot be achieved with a bimetric theory~\cite{Hohmann:2009bi} and that the acceleration~\eqref{eqn:accel} becomes positive for \(N \geq 3\) metrics~\cite{Hohmann:2010vt}. The full set of conditions~\eqref{eqn:cond} further reproduces our result that the simple action used in our first cosmological model~\cite{Hohmann:2010vt}, which is obtained from the action~\eqref{eqn:repaction} for \(c_{3} = \ldots = c_{12} = 0\), is not consistent with experiments, but consistency can be achieved for non-vanishing values of the parameters \(c_{3}, \ldots, c_{12}\), while still retaining the cosmological acceleration~\cite{Hohmann:2010ni}. In summary, we have thus confirmed the consistency of repulsive gravity models for dark energy with solar system experiments at the post-Newtonian level.

\section{Conclusion}\label{sec:conclusion}
In this article we developed an extension of the parameterized post-Newtonian (PPN) formalism to gravity theories with \(N \geq 2\) standard model copies and a corresponding number of metric tensors. Our results allow a characterization of multimetric gravity theories by a set of constant PPN parameters, in analogy to the standard PPN formalism for a single metric. We found that thirteen parameters are physical and accessible through experiments in the visible matter sector. The calculated values of these visible PPN parameters can thus be compared to their measured values obtained from high precision solar system experiments. As an illustrative example with \(N = 1\) metric we applied our formalism to general relativity and re-calculated its post-Newtonian limit. We then calculated the post-Newtonian limit of a previously discussed class of multimetric theories. Comparing our result with the measured PPN parameters we found that a subclass of these theories is compatible with all tests of post-Newtonian gravity in the solar system. This in particular applies to repulsive gravity models, in which a repulsive gravitational interaction between the different standard model copies causes an accelerating expansion of the universe.

The parameters obtained from the standard PPN formalism for theories with a single metric tensor are closely linked to physical effects, such as preferred-frame effects. We have shown that a similar interpretation is also possible for the multimetric PPN parameters. We derived conditions on the PPN parameters which indicate whether the underlying multimetric gravity theory exhibits preferred-frame effects. Relations to further physical effects will be examined in future work. Of particular interest will be the study of conservation laws in multimetric gravity. While local conservation laws, such as the covariant conservation of the energy-momentum tensor, always exist, the existence of global conservation laws, such as the conservation of total energy and momentum, depends on the gravitational background. At the post-Newtonian level their existence or non-existence can be deduced from the standard PPN parameters. For multimetric gravity we expect more general conservation laws to hold: while the total energy and momentum may be conserved, they may be transferred from one matter sector to another, leading to an apparent violation of energy and momentum conservation.

We finally remark that although we have confirmed the experimental consistency of a class of multimetric gravity theories, we have not excluded the existence of further viable theories with different PPN parameters. This is due to the fact that our formalism comprises a larger number of thirteen visible PPN parameters, compared to ten standard PPN parameters which have been measured by current experiments. A theory for which the newly introduced PPN parameters vanish, while the standard PPN parameters agree with their measured values, is thus experimentally consistent. Theories in which the three additional parameters take different values may or may not be experimentally consistent. Further calculations need to show how non-vanishing values of these parameters would influence the outcomes of current solar system tests of post-Newtonian gravity, and which bounds we can obtain from current experiments. If it turns out that some of the newly introduced parameters are still experimentally undetermined, it needs to be studied which experiments would be necessary for their measurement.

\appendix
\section{PPN parameters of multimetric repulsive gravity}\label{sec:repparam}
In this appendix we display the full set of multimetric PPN parameters for the gravity theory defined by the action~\eqref{eqn:repaction} under the experimental consistency conditions~\eqref{eqn:cond}, using the notation introduced in~\eqref{eqn:symnotation}.
\begin{align}
\bar{\alpha} &= Nz\,, & \hat{\alpha} &= 1 - z\,,\\
\bar{\gamma} &= Nz\,, & \hat{\gamma} &= 1 - z\,,\\
\bar{\theta} &= 0\,, & \hat{\theta} &= 0\,,\\
\bar{\sigma}_+ &= -2Nz\,, & \hat{\sigma}_+ &= 2(z - 1)\,,\\
\bar{\sigma}_- &= \frac{N}{2}(1 - 4z)\,, & \hat{\sigma}_- &= 2(z - 1)\,,\\
\bar{\omega}_1 &= N(5z - 1)\,, & \hat{\omega}_1 &= 5(1 - z)\,,\\
\bar{\omega}_2 &= N(1 - z)\,, & \hat{\omega}_2 &= z - 1\,,\\
\bar{\phi}_p &= 2N(4z - 1)\,, & \hat{\phi}_p &= 8(1 - z)\,,\\
\bar{\phi}_{\Pi} &= 2Nz\,, & \hat{\phi}_{\Pi} &= 2(1 - z)\,,
\end{align}
\begin{subequations}
\begin{align}
\bar{\psi}_1 &= \frac{N(z - 1)^3(3N - 4 - (N - 4)z)}{2(z + N - 1)}(c_{6} - c_{8} - 2c_{4}) - \frac{19N^2(z - 1)^3(z + 1)}{2(z + N - 1)}c_{10}\nonumber\\
&\phantom{=}+ \frac{N(z - 1)^2(N + 1 + (2N - 1)z)}{2(z + N - 1)}\,,\\
\ola{\psi}_1 &= \frac{(2N - 1)(z - 1)^3(3N - 4 - (N - 4)z)}{6(z + N - 1)}(2c_{4} - c_{6} + c_{8}) + \frac{19N(2N - 1)(z - 1)^3(z + 1)}{6(z + N - 1)}c_{10}\nonumber\\
&\phantom{=}+ 2(z - 1)^2(Nz - z + 1)(c_{11} + 2c_{12})\nonumber\\
&\phantom{=}+ \frac{(z - 1)(2N^2 + 3N - 3 + (8N^2 - 13N + 6)z - (4N^2 - 10N + 3)z^2)}{6(z + N - 1)}\,,\\
\ora{\psi}_1 &= \frac{(z - 1)^3(3N - 4 - (N - 4)z)}{2(z + N - 1)}(2c_{4} - c_{6} + c_{8}) + \frac{19N(z - 1)^3(z + 1)}{2(z + N - 1)}c_{10}\nonumber\\
&\phantom{=}- \frac{(z - 1)^2(N + 1 + (2N - 1)z)}{2(z + N - 1)}\,,\\
\tilde{\psi}_1 &= \frac{(N + 1)(z - 1)^3(3N - 4 - (N - 4)z)}{6(z + N - 1)}(2c_{4} - c_{6} + c_{8}) + \frac{19N(N + 1)(z - 1)^3(z + 1)}{6(z + N - 1)}c_{10}\nonumber\\
&\phantom{=}- 2(z - 1)^2(Nz - z + 1)(c_{11} + 2c_{12})\nonumber\\
&\phantom{=}+ \frac{(z - 1)(N^2 + 3 - (5N^2 - 7N + 6)z - (2N^2 + 7N - 3)z^2)}{6(z + N - 1)}\,,\\
\hat{\psi}_1 &= \frac{(z - 1)^3(3N - 4 - (N - 4)z)}{2(z + N - 1)}(c_{6} - c_{8} - 2c_{4}) - \frac{19N(z - 1)^3(z + 1)}{2(z + N - 1)}c_{10}\nonumber\\
&\phantom{=}+ \frac{(z - 1)^2(N + 1 + (2N - 1)z)}{2(z + N - 1)}\,,
\end{align}
\end{subequations}
\begin{align}
\bar{\psi}_2 &= 0\,, & \ola{\psi}_2 &= 0\,, & \ora{\psi}_2 &= 0\,, & \tilde{\psi}_2 &= 0\,, & \hat{\psi}_2 &= 0\,,
\end{align}
\begin{subequations}
\begin{align}
\bar{\psi}_3 &= \frac{N(z - 1)^3(15N - 22 - (7N - 22)z)}{4(z + N - 1)}(c_{6} - c_{8} - 2c_{4}) - \frac{19N(z - 1)^3(5N + 2 + (7N - 2)z)}{4(z + N - 1)}c_{10}\nonumber\\
&\phantom{=}+ N\frac{5N + 9 + (2N^2 + 2N - 27)z - (10N^2 + 11N - 27)z^2 + (4N - 9)z^3}{4(z + N - 1)}\,,\\
\ola{\psi}_3 &= \frac{(z - 1)^3(24N^2 - 35N - 2 - (8N^2 - 27N - 2)z)}{12(z + N - 1)}(2c_{4} - c_{6} + c_{8})\nonumber\\
&\phantom{=}- \frac{(z - 1)^2(152N^2 + 29N + 66 + (48N^2 + 18N - 132)z - (152N^2 + 47N - 66)z^2)}{12(z + N - 1)}c_{10}\nonumber\\
&\phantom{=}+ 4(z - 1)^2(Nz - z + 1)(c_{11} + 2c_{12})\nonumber\\
&\phantom{=}+ \frac{(z - 1)(8N^2 + 15N + 3 + (62N^2 - 67N - 6)z - (16N^2 - 52N - 3)z^2}{12(z + N - 1)}\,,\\
\ora{\psi}_3 &= \frac{(z - 1)^3(15N - 22 - (7N - 22)z)}{4(z + N - 1)}(2c_{4} - c_{6} + c_{8})\nonumber\\
&\phantom{=}- \frac{(z - 1)^2(111N + 22 + (16N^2 + 6N - 44)z - (117N - 22)z^2)}{4(z + N - 1)}c_{10}\nonumber\\
&\phantom{=}+ \frac{(z - 1)(5N + 9 + (8N^2 + N - 18)z - (6N - 9)z^2)}{4(z + N - 1)}\,,\\
\tilde{\psi}_3 &= \frac{(z - 1)^3(21N^2 - 31N + 2 - (13N^2 - 39N + 2)z)}{12(z + n - 1)}(2c_{4} - c_{6} + c_{8})\nonumber\\
&\phantom{=}- \frac{(z - 1)^2(133N^2 + 37N - 18 + (18N^2 - 150N + 36)z - (247N^2 - 113N + 18)z^2}{12(z + n - 1)}c_{10}\nonumber\\
&\phantom{=}- 4(z - 1)^2(Nz - z + 1)(c_{11} + 2c_{12})\nonumber\\
&\phantom{=}+ \frac{(z - 1)(N^2 + 18N - 3 + (N^2 - 29N + 6)z - (26N^2 - 11N + 3)z^2)}{12(z + n - 1)}\,,\\
\hat{\psi}_3 &= \frac{(z - 1)^3(15N - 22 - (7N - 22)z)}{4(z + N - 1)}(c_{6} - c_{8} - 2c_{4}) - \frac{19(z - 1)^3(5N + 2 + (7N - 2)z)}{4(z + N - 1)}c_{10}\nonumber\\
&\phantom{=}- \frac{(z - 1)^2(5N - 19 - (14N - 19)z)}{4(z + N - 1)}\,,
\end{align}
\end{subequations}
\begin{align}
\bar{\psi}_4 &= 0\,, & \ola{\psi}_4 &= 0\,, & \ora{\psi}_4 &= 0\,, & \tilde{\psi}_4 &= 0\,, & \hat{\psi}_4 &= 0\,,
\end{align}
\begin{subequations}
\begin{align}
\bar{\psi}_5 &= \frac{N(z - 1)^3(3N - 4 - (N - 4)z)}{2(z + N - 1)}(c_{6} - c_{8} - 2c_{4}) - \frac{19N^2(z - 1)^3(z + 1)}{2(z + N - 1)}c_{10}\nonumber\\
&\phantom{=}+ \frac{N(z - 1)(z - N - 1)(Nz - z + 1)}{2(z + N - 1)}\,,\\
\ola{\psi}_5 &= \frac{(2N - 1)(z - 1)^3(3N - 4 - (N - 4)z)}{6(z + N - 1)}(2c_{4} - c_{6} + c_{8}) + \frac{19N(2N - 1)(z - 1)^3(z + 1)}{6(z + N - 1)}c_{10}\nonumber\\
&\phantom{=}+ 2(z - 1)^2(Nz - z + 1)(c_{11} + 2c_{12})\nonumber\\
&\phantom{=}+ \frac{(z - 1)(2N^2 + 3N - 3 + (8N^2 - 13N + 6)z - (4N^2 - 10N + 3)z^2)}{6(z + N - 1)}\,,\\
\ora{\psi}_5 &= \frac{(z - 1)^3(3N - 4 - (N - 4)z)}{2(z + N - 1)}(2c_{4} - c_{6} + c_{8}) + \frac{19N(z - 1)^3(z + 1)}{2(z + N - 1)}c_{10}\nonumber\\
&\phantom{=}- \frac{(z - 1)(z - N - 1)(Nz - z + 1)}{2(z + N - 1)}\,,\\
\tilde{\psi}_5 &= \frac{(N + 1)(z - 1)^3(3N - 4 - (N - 4)z)}{6(z + N - 1)}(2c_{4} - c_{6} + c_{8}) + \frac{19N(N + 1)(z - 1)^3(z + 1)}{6(z + N - 1)}c_{10}\nonumber\\
&\phantom{=}- 2(z - 1)^2(Nz - z + 1)(c_{11} + 2c_{12})\nonumber\\
&\phantom{=}- \frac{(z - 1)(2N^2 - 3N - 3 + (2N^2 - N + 6)z + (2N^2 + 4N - 3)z^2)}{6(z + N - 1)}\,,\\
\hat{\psi}_5 &= \frac{(z - 1)^3(3N - 4 - (N - 4)z)}{2(z + N - 1)}(c_{6} - c_{8} - 2c_{4}) - \frac{19N(z - 1)^3(z + 1)}{2(z + N - 1)}c_{10}\nonumber\\
&\phantom{=}+ \frac{(z - 1)^2(Nz - z + 1)}{z + N - 1}\,,
\end{align}
\end{subequations}
\begin{subequations}
\begin{align}
\bar{\psi}_6 &= \frac{3N(N - 2)(z - 1)^4}{4(z + N - 1)}(2c_{4} - c_{6} + c_{8}) - \frac{19N(z - 1)^3(N + 2 + (3N - 2)z)}{4(z + N - 1)}c_{10}\nonumber\\
&\phantom{=}+ N\frac{N + 5 + (4N - 15)z - (8N^2 + 3N - 15)z^2 - (2N + 5)z^3}{4(z + N - 1)}\,,\\
\ola{\psi}_6 &= -\frac{(z - 1)^2(12N^2 - 70N + 62 - (55N^2 - 142N + 124)z + (13N^2 - 72N + 62)z^2)}{6(z + N - 1)}c_{4}\nonumber\\
&\phantom{=}+ \frac{(z - 1)^2(12N^2 - 28N + 20 - (13N^2 - 58N + 40)z + (13N^2 - 30N + 20)z^2)}{12(z + N - 1)}c_{6}\nonumber\\
&\phantom{=}- \frac{(z - 1)^2(12N^2 - 76N + 68 - (61N^2 - 154N + 136)z + (13N^2 - 78N + 68)z^2)}{12(z + N - 1)}c_{8}\nonumber\\
&\phantom{=}- \frac{(z - 1)^2(76N^2 - 200N + 48 - (105N^2 + 18N + 96)z - (247N^2 - 218N - 48)z^2)}{12(z + N - 1)}c_{10}\nonumber\\
&\phantom{=}+ 14(z - 1)^2(Nz - z + 1)(c_{11} + c_{12})\nonumber\\
&\phantom{=}+ \frac{(z - 1)(4N^2 + 36N - 48 + (49N^2 - 119N + 96)z - (26N^2 - 83N + 48)z^2)}{12(z + N - 1)}\,,\\
\ora{\psi}_6 &= \frac{(z - 1)^2(12N^2 - 88N + 98 - (55N^2 - 178N + 196)z + (13N^2 - 90N + 98)z^2)}{6(z + N - 1)}c_{4}\nonumber\\
&\phantom{=}- \frac{(z - 1)^2(12N^2 - 46N + 56 - (13N^2 - 94N + 112)z + (13N^2 - 48N + 56)z^2)}{12(z + N - 1)}c_{6}\nonumber\\
&\phantom{=}+ \frac{(z - 1)^2(12N^2 - 94N + 104 - (61N^2 - 190N + 208)z + (13N^2 - 96N + 104)z^2)}{12(z + N - 1)}c_{8}\nonumber\\
&\phantom{=}+ \frac{(z - 1)^2(76N^2 - 410N - 84 - (201N^2 + 54N - 168)z - (247N^2 - 464N + 84)z^2)}{12(z + N - 1)}c_{10}\nonumber\\
&\phantom{=}- 14(z - 1)^2(Nz - z + 1)(c_{11} + c_{12})\nonumber\\
&\phantom{=}- \frac{(z - 1)(4N^2 + 30N - 78 + (N^2 - 101N + 156)z - (26N^2 - 71N + 78)z^2)}{12(z + N - 1)}\,,\\
\tilde{\psi}_6 &= -\frac{(z - 1)^2(19N^2 - 13N - 6 + (6N^2 - 50N + 12)z - (57N^2 - 63N + 6)z^2)}{4(z + N - 1)}c_{10}\nonumber\\
&\phantom{=}+ \frac{3(N - 1)(N - 2)(z - 1)^4}{4(z + N - 1)}(c_{6} - c_{8} - 2c_{4}) - \frac{(N - 1)(z - 1)^2(N + 5 + (6N - 5)z)}{4(z + N - 1)}\,,\\
\hat{\psi}_6 &= \frac{3(N - 2)(z - 1)^4}{4(z + N - 1)}(2c_{4} - c_{6} + c_{8}) - \frac{19(z - 1)^3(N + 2 + (3N - 2)z)}{4(z + N - 1)}c_{10}\nonumber\\
&\phantom{=}- \frac{(z - 1)^2(7N - 13 - (6N - 13)z)}{4(z + N - 1)}\,,
\end{align}
\end{subequations}
\begin{align}
\bar{\psi}_7 &= 0\,, & \ola{\psi}_7 &= 0\,, & \ora{\psi}_7 &= 0\,, & \tilde{\psi}_7 &= 0\,, & \hat{\psi}_7 &= 0\,.
\end{align}

\acknowledgments
The author is happy to thank Laur J\"arv, Piret Kuusk, Hannes Liivat and Erik Randla for helpful discussions and support. He gratefully acknowledges full financial support from the Estonian Research Council through the Postdoctoral Research Grant ERMOS115.

\end{document}